%% file: main.tex
\documentclass[aps,prx,twocolumn,notitlepage,showkeys,showpacs,superscriptaddress,longbibliography,footinbib, preprintnumbers]{revtex4-2}
\usepackage{tikz-cd}
\usepackage{tikz}
\usepackage{cancel}
\usepackage{placeins}
\usepackage{float}
\usetikzlibrary{graphs,quotes}
\usepackage{mhchem}

\usepackage{array}

\input{./preamble}

\begin{document}
\title{Mirror-Selective Quasiparticle Interference in Bilayer Nickelate Superconductor}

\author{Zhongyi Zhang}
\thanks{Corresponding authors:~\href{mailto:zhongyizhang@ust.hk}{zhongyizhang@ust.hk}}
\affiliation{Department of Physics, Hong Kong University of Science and Technology, Clear Water Bay, Hong Kong, China}

\author{Jun Zhan}
\affiliation{Beijing National Laboratory for Condensed Matter Physics and Institute of Physics, Chinese Academy of Sciences, Beijing 100190, China}
\affiliation{School of Physical Sciences, University of Chinese Academy of Sciences, Beijing 100190, China}

\author{Congcong Le}
\affiliation{RIKEN Interdisciplinary Theoretical and Mathematical Sciences (iTHEMS), Wako, Saitama 351-0198, Japan}

\author{Hoi Chun Po}
\affiliation{Department of Physics, Hong Kong University of Science and Technology, Clear Water Bay, Hong Kong, China}

\author{Jiangping Hu}
\thanks{Corresponding authors:~\href{mailto:jphu@iphy.ac.cn}{jphu@iphy.ac.cn}}
\affiliation{Beijing National Laboratory for Condensed Matter Physics and Institute of Physics, Chinese Academy of Sciences, Beijing 100190, China}

\author{Xianxin Wu}
\thanks{Corresponding authors:~\href{mailto:xxwu@itp.ac.cn}{xxwu@itp.ac.cn}}
\affiliation{CAS Key Laboratory of Theoretical Physics, Institute of Theoretical Physics, Chinese Academy of Sciences, Beijing, China}

\begin{abstract}

The recent discovery of high-temperature superconductivity in both bulk and thin-film bilayer nickelates has garnered significant attention. Despite extensive research, the electronic structure essential for superconductivity and pairing symmetry remain elusive, fueling debates over the underlying pairing mechanism. In this study, inspired by recent STM experiments on thin films, we investigate the quasiparticle interference (QPI) characteristics of bilayer nickelates in both normal and superconducting states to identify their Fermiology and pairing symmetry. 
We demonstrate that the mirror symmetry inherent in the bilayer structure induces mirror-selective quasiparticle scattering by establishing selection rules based on the mirror properties of impurities and the mirror eigenvalues of electronic wavefunctions.
This mirror-selective scattering allows for the differentiation of distinct Fermiologies, as QPI patterns vary markedly between scenarios with and without the $d_{z^2}$-bonding Fermi surface (FS).
Furthermore, it enables the separate detection of sign changes in superconducting gaps both within the same FS and between different FSs.
Crucially, if the mirror-symmetry-enforced selection rules are ignored, the QPI response of an $s_\pm$-wave state can masquerade as that of a conventional $s$-wave state, leading to a misidentification of the pairing symmetry.
When combined with field-dependent and reference QPI measurements, this approach facilitates the precise determination of pairing symmetry, even in the presence of FS-dependent gaps and gap anisotropy.
Additionally, we discuss practical considerations for STM measurements to effectively identify the pairing symmetry. Our findings demonstrate that mirror-selective QPI is a powerful tool for distinguishing between different Fermiologies and pairing states, which is helpful in pinning down pairing symmetry and revealing the pairing mechanism in bilayer nickelates.

\end{abstract}
\maketitle

The recent discovery of high-temperature $T_c$ superconductivity in pressurized bilayer nickelate La$_3$Ni$_2$O$_7$ (LNO) in the Ruddlesden-Popper phase has generated widespread interests~\cite{sun2023}, establishing the nickelate family as the third high-$T_c$ family alongside cuprates and iron based superconductors. In contrast to their predecessors, the strongly coupled bilayer structure through apical oxygens produces a $d^{7.5}$ configuration in the Ni$^{2.5+}$, with both $d_{x^2-y^2}$ and $d_{z^2}$ orbitals contributing to the low-energy electronic structure~\cite{YaoDX,YZhang2023,Lechermann2023,Hirofumi2023possible,XWu,XJZhou2023,HHWen2023}. Recently, compressively strained LNO thin films grown on the SrLaAlO$_4$ substrate have been found to exhibit superconductivity with a $T_c$ exceeding 40 K at ambient pressure ~\cite{ko2024signatures,zhou2024ambientpressuresuperconductivityonset40,liu2025superconductivitynormalstatetransportcompressively,bhatt2025resolvingstructuraloriginssuperconductivity}. 
Despite intensive studies, the electronic structure relevant for high $T_c$ superconductivity and the pairing mechanism remain elusive. 
In the bulk, the Lifshitz transition under pressure that the $d_{z^2}$ interlayer bonding state crosses the Fermi level $E_{\text{F}}$ is crucial for bulk superconductivity according to theoretical calculations~\cite{sun2023}. However, direct experimental verification under pressure remains challenging. Intriguingly, recent ARPES measurements from two groups reveal conflicting Fermiologies in superconducting thin films~\cite{ChenZY_ARPES2025,ShenZX2025}: one group report the $d_{z^2}$ bonding state crossing $E_{\text{F}}$, while another group find it located below $E_{\text{F}}$. Regarding the pairing mechanism and symmetry, no theoretical consensus has been reached~\cite{Wang327prb, lu2024interlayer,HYZhangtype2,XWu,FangYang327prl,WeiLi327prl,Hirofumi2023possible,YifengYang327prb,YifengYang327prb2,YiZhuangYouSMG,tian2023correlation,Dagotto327prb,zhang2024structural,Jiang_2024,PhysRevB.108.L201121,liao2023electron,ryee2024quenched,luo2023hightc,KJiang:17402,KuWeiprl,fan2023superconductivity,zhan2024cooperation,ChenHH2025,PhysRevB.111.144514}. Spin fluctuations are believed to primarily promote an $s_{\pm}$-wave pairing~\cite{Hirofumi2023possible,XWu,Wang327prb,FangYang327prl}, and also $d_{xy}$-wave pairing with small crystal field splitting~\cite{Lechermann2023,ChenHH2025}. In the strong-coupling regime, interlayer exchange coupling, Hund's rule coupling, and orbital hybridization are emphasized in generating interlayer $s$-wave pairing~\cite{lu2024interlayer,HYZhangtype2,WeiLi327prl}, while other studies suggest spin-orbital exchange coupling contributes to $d_{x^2-y^2}$-wave pairing~\cite{KJiang:17402,KuWeiprl,fan2023superconductivity}. Additionally, the interlayer interorbital $d$-wave pairing is favored when the interlayer repulsion is significant~\cite{LiJX_nonlocal,2025arXiv250318877Z}. 

Identifying Fermiology and pairing symmetry plays a crucial role in understanding the underlying pairing mechanism~\cite{RevModPhys.84.1383}. Several proposals have been proposed in the bulk LNO systems~\cite{jiangarxiv2025,Ilya_cp2025,PhysRevB.111.174525,PhysRevB.109.L180502,yang2025possible}. Thanks to superconducting thin films at ambient pressure, several powerful experimental techniques measurements such as angle-resolved photoemission spectroscopy (ARPES) and scanning tunneling microscopy (STM) have been employed~\cite{ChenZY_ARPES2025,2025arXiv250217831S,ShenZX2025,2025arXiv250707409S,WenHH_STM2025}.
Fourier-transform STM, known as quasiparticle interference (QPI), is especially promising, as it can not only provide insights into Fermiology but also identify sign-changing structures of superconducting gaps-an essential signature for determining pairing symmetry, as demonstrated in cuprates~\cite{doi:10.1126/science.1072640,McElroy2003,hanaguri2009coherence,Kohsaka2008} and iron-based superconductors~\cite{Allan2013,hanaguri2010unconventional}. 
The sign-changing structure is inferred from the enhancement of scattering at certain $\qqq$ vectors depending on the time-reversal nature of impurity potential. Coherence factors embedded in the quasiparticle wavefunctions mean time-reversal symmetry dictates selection rules for scattering enhancement, revealing information about sign-changed gaps. Within the bilayer structure, the presence of mirror reflection symmetry will significantly impact QPI and introduce additional selection rules in both normal and superconducting states, which can help to resolve the ongoing debate on Fermiolgy and pairing symmetry.

Motivated by these developments, we study QPI characteristics of bilayer nickelates in both normal and superconducting states to identify their Fermiology and pairing symmetry. We demonstrate that the mirror symmetry inherent in the bilayer structure dictates mirror-selective QPI patterns by introducing selection rules based on the mirror properties of impurities and the mirror eigenvalues of the wavefunctions. 
In the normal state, mirror-even impurity potentials exclusively facilitate scattering between bonding (or antibonding) Fermi surfaces (FSs) with identical mirror eigenvalues, whereas mirror-odd impurity potentials enable scattering between bonding and antibonding FSs with opposite mirror eigenvalues. This results in distinctly different QPI patterns depending on the presence of the $d_{z^2}$-bonding pocket, thereby allowing experimental differentiation of the Fermiology.
In the superconducting state, mirror-selective scattering allows for the separate detection of sign changes in the superconducting gaps both within the same FS and between different FSs.
Importantly, if the mirror-symmetry-enforced selection rules are neglected, the QPI signatures of an $s_{\pm}$-wave state can appear indistinguishable from those of a conventional $s$-wave state, leading to a misidentification of the pairing symmetry.
When combined with field-dependent and reference QPI measurements, this approach enables clear differentiation between $s$-, $s_{\pm}$- and $d_{x^2-y^2}$-wave pairing symmetries, even in the presence of FS-dependent gaps and gap anisotropy. 
Finally, we discuss practical considerations for STM measurements to effectively identify the pairing symmetry in bilayer nickelates.

\begin{figure}[t]
	\centering
	\includegraphics[width=0.9875\linewidth]{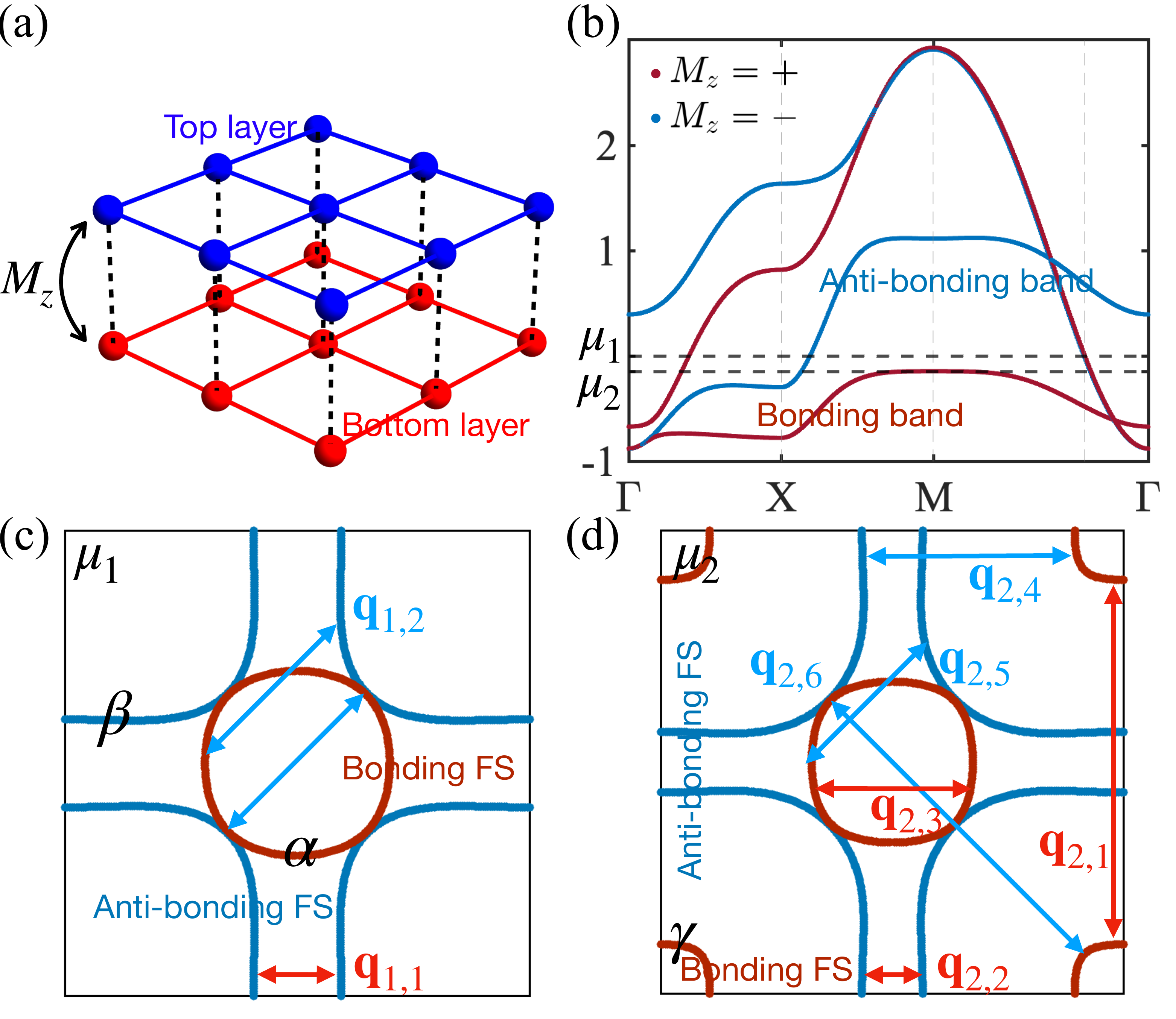}
	\caption{\label{main_fig_bilayer_structure} {\bf Bilayer structure and Fermi surface for bilayer nickelates.}
	(a) The mirror-symmetric bilayer structure. Each site hosts two $3d$ orbitals $d_{x^2-y^2}$ and $d_{z^2}$ of the nickel atom.
	(b) The band structure of the Hamiltonian described in Eq.~\eqref{main_eq_Hamiltonian}.
	$\mu_1$ and $\mu_2$ represent pristine filling $n=3.0$ and filling $n=2.7$, respectively.
	(c)-(d) Two types of Fermiologies with $\mu_1=0$ and $\mu_2=-0.1433$.
	The corresponding vector $\qqq_{i,j}$ means $j$-th vectors at Fermi level $\mu_i$.
	In (b)-(d), red and blue mean the states are the bonding and antibonding states.
	From the $\Gamma$ to the M, the Fermi surfaces are labeled as $\alpha$, $\beta$, and $\gamma$, respectively.
	Red arrows and Blue arrows represent vectors connecting FS segments with same and different mirror eigenvalues.
	}
\end{figure}

\textit{Effective model and impurity scattering on the bilayers.--}
We start with the two-orbital effective tight-binding model of bilayer nickelates, as shown in Fig.~\ref{main_fig_bilayer_structure}(a).
Each lattice site hosts two orbitals, which are the $3d$ orbitals $d_{x^2-y^2}$ and $d_{z^2}$ of the nickel atom.
The tight-binding Hamiltonian can be written as
\begin{equation}\label{main_eq_Hamiltonian}
    \hat{\mathcal{H}}_0(\kk)=\sum_{\kk,\ell\ell^\prime,oo^\prime}\hat{c}^\dagger_{\kk,\ell o}H_{\ell\ell^\prime,oo^\prime}(\kk)\hat{c}_{\kk,\ell^\prime o^\prime},
\end{equation}
where the $\hat{c}_{\kk,\ell,o}(\hat{c}^\dagger_{\kk,\ell,o})$ is the annihilation (creation) operator of Ni 3$d$ electron with orbital $o$ and momentum $\kk$ in the layer $\ell$.
The matrix elements of $H_{\ell\ell^\prime,oo^\prime}(\kk)$ are given in Ref.~\cite{le2025landscape}, and the corresponding band structure is shown in Fig.~\ref{main_fig_bilayer_structure}(b).
In the following, we use $\ell_i$ and $\sigma_j$ to denote the Pauli matrices in the layer and orbital spaces, respectively. As the $d_{z^2}$ bonding state's crossing of $E_{\text{F}}$ in thin films is still under debate, we adjust the chemical potential to generate two Fermiologies: (1) with $\mu=\mu_1$, the bonding $d_{z^2}$ state is below $E_{\text{F}}$, resulting in $\alpha$ and $\beta$ pockets, as shown in Fig.~\ref{main_fig_bilayer_structure}(c); (2) with $\mu=\mu_2$, the bonding $d_{z^2}$ state crosses  $E_{\text{F}}$ and an additional $\gamma$ pocket around the corner appears, 
creating three pockets, as shown in Fig.~\ref{main_fig_bilayer_structure}(d).

In the high-symmetry phase of bilayer nickelates, such a bilayer structure possesses a mirror symmetry that relates the top and bottom layers, whose representation is $\mathcal{M}_z=\ell_x\sigma_0$.
As a result, each FS corresponds to a symmetric (bonding) or antisymmetric (antibonding) combination of wavefunctions from the top and bottom layers, carrying mirror eigenvalues of $+1$ and $-1$, respectively. The bonding $\alpha,\gamma$ FSs are mirror-even, while the anti-binding $\beta$ is mirror-odd.
According to this mirror symmetry and impurity locations, there are three representative types of naturally occurring nonmagnetic impurities with the bilayer structure.
Two type of intralayer impurities are mirror-even impurity $V_1$ and mirror-odd impurity $V_2$, defined as,
\begin{equation}
	V_{1/2}=v_0\ell_{0/z}\sigma_0,
\end{equation}
with $v_0$ being the strength of impurity.
The third type consists of impurities situated between two NiO$_2$ layers, such as the inner apical oxygen.
Owing to the distinct spatial orientation of $d_{x^2-y^2}$ and $d_{z^2}$ orbitals, this impurity scattering potential mainly occurs on the $d_{z^2}$ orbital, exhibiting orbital-selective behavior.
Therefore, we adopt the following mirror-even form for the interlayer impurity:
\begin{equation}
	V_3=\frac{v_0}{2}\ell_x(\sigma_0-\sigma_z).
\end{equation}

\begin{table*}
\centering
\caption{{\bf Selection rules for different types of impurity scattering in bilayer nickelates.}
BFS and ABFS denote the bonding ($\alpha,\gamma$) and antibonding ($\beta$) Fermi surfaces, respectively.
BFS $\leftrightarrow$ BFS means the scattering occurs between BFS or between ABFS, whereas BFS $\leftrightarrow$ ABFS means the scattering occurs between BFS and ABFS.
$\Theta$ is the time-reversal symmetry and $\phi_\kk$ is the phase of the superconducting gap $\Delta(\kk)=|\Delta(\kk)|e^{i\phi_\kk}$.
The second-to-last column indicates how each type of impurity couples to states with different mirror eigenstates, where $\Leftrightarrow$ represents the transformation of BFS (ABFS) into ABFS (BFS).
The last column corresponds to the symmetries preserved by the renormalized $T$-matrix.
\label{main_table_scattering_matrix}}
\begin{tabular}{ p{3cm}<{\centering}|p{3.3cm}<{\centering}|p{1cm}<{\centering}|p{2.5cm}<{\centering}|p{0.9cm}<{\centering}|p{2cm}<{\centering}|p{1.8cm}<{\centering}|p{1.8cm}<{\centering} }
\hline\hline
Scattering potential $V_i$    & Realization      & $\mathcal{M}_z$ & $\mathcal{M}_z$-selection & $\Theta$& $\Theta$-selection & BFS/ABFS & $T(\omega)$  \\\hline
$ \tau_z\ell_0\sigma_0$  &Nonmag. top $+$ bottom  &even & BFS $\leftrightarrow$ BFS & \multirow{3}*{even} & $\phi_\kk=-\phi_{\kk+\qqq}$ & $+/+$& $\mathcal{M}_z$ \\ \cline{1-4} \cline{6-8}
$ \tau_z\ell_z\sigma_0$  & Nonmag. top $-$ bottom  &odd & BFS $\leftrightarrow$ ABFS & &  $\phi_\kk=-\phi_{\kk+\qqq}$ & $\Longleftrightarrow$ & $-$ \\ \cline{1-4} \cline{6-8}
$ \tau_z\ell_x(\sigma_0-\sigma_z)$  &  Nomagn. interlayer &even&BFS $\leftrightarrow$ BFS & &$\phi_\kk=-\phi_{\kk+\qqq}$ & $+/-$ & $\mathcal{M}_z$ \\ \hline
$ \tau_0\ell_0\sigma_0$  &Mag. top $+$ bottom  &even & BFS $\leftrightarrow$ BFS&\multirow{3}*{odd} &$\phi_\kk=\phi_{\kk+\qqq}$ & $+/+$ & $\mathcal{M}_z$ \\ \cline{1-4} \cline{6-8}
$ \tau_0\ell_z\sigma_0$  &Mag. top $-$ bottom  &odd & BFS $\leftrightarrow$ ABFS& &$\phi_\kk=\phi_{\kk+\qqq}$ & $\Longleftrightarrow$&$-$ \\ \cline{1-4} \cline{6-8}
$ \tau_0\ell_x(\sigma_0-\sigma_z)$  &Mag. interlayer &even & BFS $\leftrightarrow$ BFS& &$\phi_\kk=\phi_{\kk+\qqq}$ & $+/-$& $\mathcal{M}_z$ \\ 
\hline\hline
\end{tabular}
\end{table*}

\textit{Mirror-selective QPI in normal state.--}
Next, we show how mirror symmetry influences quasiparticle scattering for different impurity scattering in the normal state.
Based on the $T$-matrix approximation~\cite{RevModPhys.78.373}, the Green's function in the presence of a single impurity $\hat{\mathcal{V}}=\sum_{\ell\ell^\prime,oo^\prime}V_{\ell\ell^\prime,oo^\prime} \hat{c}^\dagger_{\kk,\ell o}\hat{c}_{\kk,\ell^\prime o^\prime}$ is given by
\begin{eqnarray}
    G(\kk,\kk+\qqq,\omega)&=&G_0(\kk,\omega)\delta_{\qqq,0}+G_0(\kk,\omega)T(\omega)G_0(\kk+\qqq,\omega) \nonumber\\
    &=&G_0(\kk,\omega)\delta_{\qqq,0}+\delta G(\kk,\kk+\qqq,\omega)
\end{eqnarray}
where the bare Green function $G_0(\kk,\omega)=[\omega-H(\kk)+i\eta]^{-1}$, and the $T$-matrix accounting for multiple scattering off the impurity is given by
\begin{equation}
\begin{split}
    T(\omega)&=V(1-\frac{1}{N}\sum_{\kk}G_{0}(\kk,\omega)V)^{-1}.
\label{Tmatrixpotential}
\end{split}
\end{equation}
with $N$ being volume of the system.
The Fourier transformation (FT) of the change of the local density of states (LDOS) $\delta\rho(\rr,\omega)$, referred to FT-QPI pattern, is given by
\begin{equation}\label{main_eq_normalQPI}
\begin{split}
    \delta\rho(\qqq,\omega)&=\frac{1}{N}\sum_{\rr}e^{-i\qqq\cdot\rr}\delta\rho(\rr,\omega)\\
    &=-\frac{1}{N\pi}\Im\text{Tr}[\sum_{\kk}\delta G(\kk,\kk+\qqq,\omega)].
\end{split}
\end{equation}

\begin{figure}[b]
	\centering
	\includegraphics[width=0.9875\linewidth]{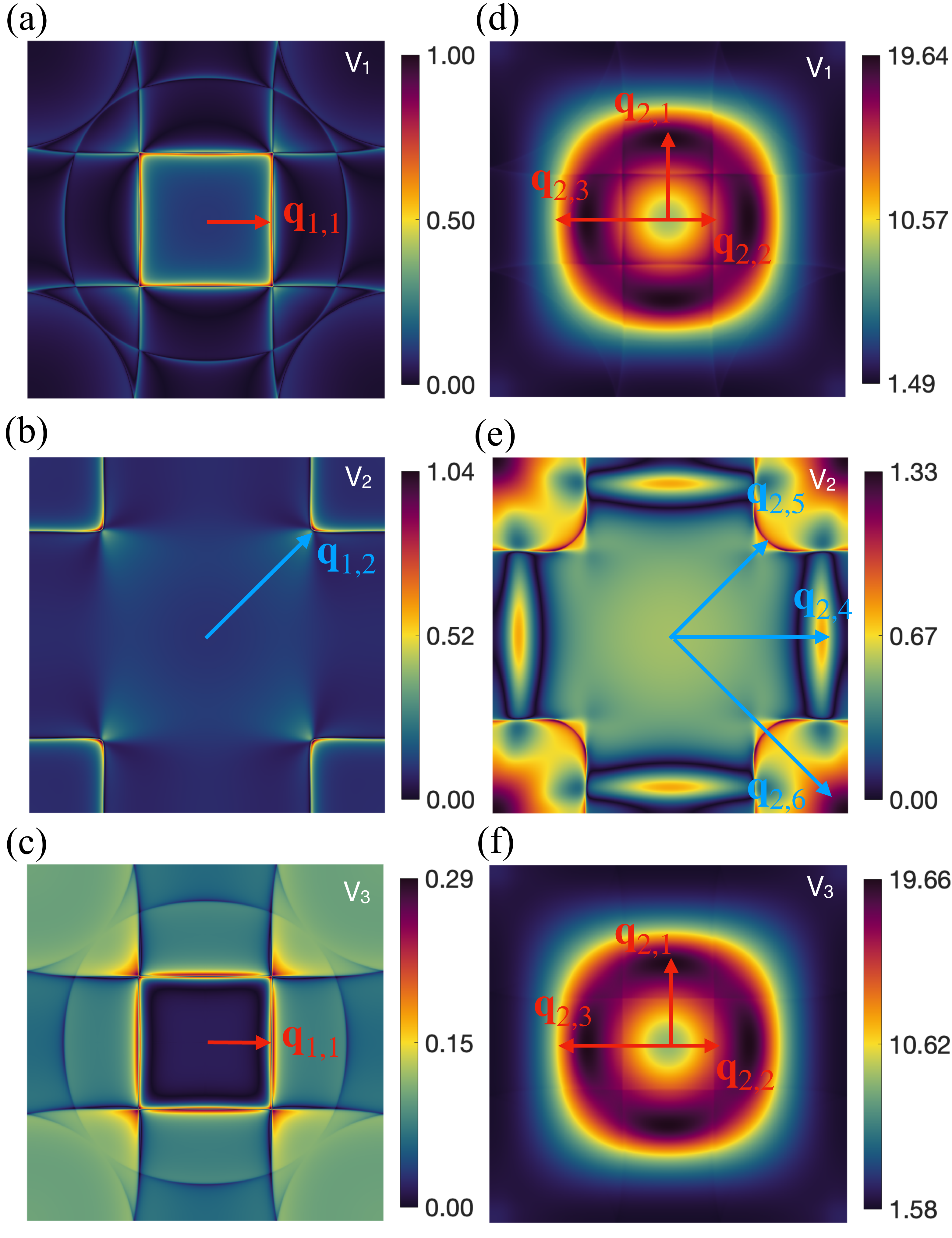}
	\caption{\label{main_fig_normal_QPI} {\bf FT-QPI intensity $|\delta\rho(\qqq,\omega=0)|$ in Eq.~\eqref{main_eq_normalQPI} of normal state for three types of impurities.}
	(a)-(c) QPI patterns for $V_1$, $V_2$, and $V_3$ without the $\gamma$-FS.
	For mirror-odd impurity $V_2$, we calculate the layer-dependent QPI.
	(d)-(f) Same quantities as (a)-(c), but with the $\gamma$-FS.
	In calculation, we adopt a $1001\times 1001$ lattice sites and $v_0=0.25$.
	}
\end{figure}

\begin{figure*}[htbp!]
	\centering
	\includegraphics[width=0.97\linewidth]{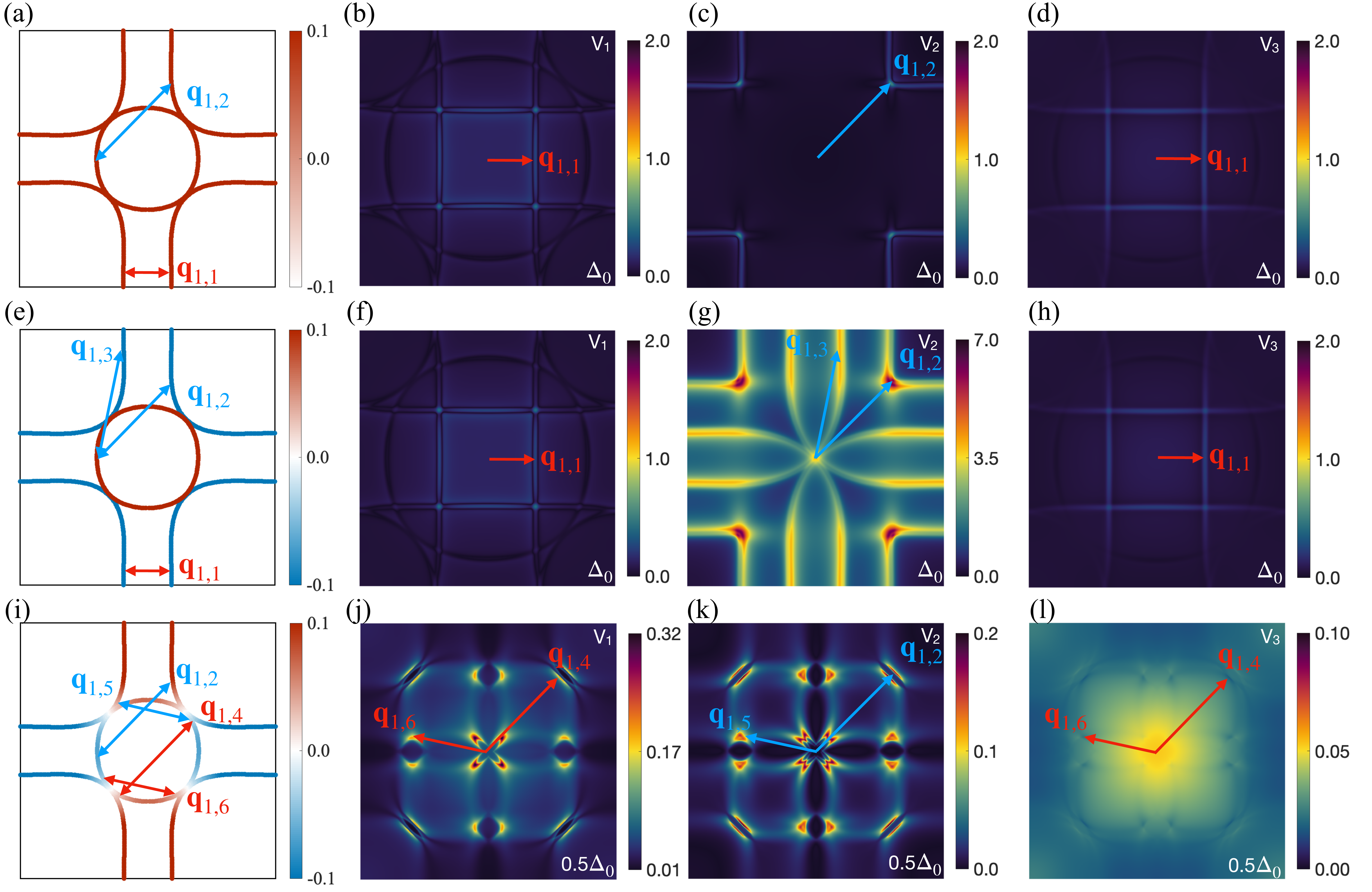}
	\caption{\label{main_fig_wog_BdG_QPI} 
	{\bf QPI intensity for three types of pairing symmetries and nonmagnetic impurities without the $\gamma$-FS.}
	(a) The $s$-wave gap amplitude on the Fermi surface.
	(b)-(d) the QPI intensity $|\delta\rho^{\text{odd}}(\qqq,\omega)|$ for three types impurities measured at the superconducting gap $\Delta_0$.
	(e)-(h) and (i)-(l) are the same quantities as (b)-(d), but for $s_\pm$-wave and $d_{x^2-y^2}$-wave.
	(f)-(h) are measured at $\omega=\Delta_0$, and (i)-(l) are measured at $\omega=\Delta_0/2$.
	The strength of impurity is set to be $v_0=0.25$.
	}
\end{figure*}
To demonstrate QPI characteristics of different impurity potentials, we consider the weak (Born) impurity scattering by taking $T(\omega)$ as the bare scattering potential. The QPI intensity $|\delta\rho(\qqq,\omega)|$ is shown in Fig.~\ref{main_fig_normal_QPI} for both cases with and without the $\gamma$-FS.
For mirror-even types of scattering potential $V_1$ and $V_3$, only those $\qqq$-vectors that connect same FS or FSs with the same mirror eigenvalue can give rise to finite peaks~\cite{PhysRevB.88.125141}. This is because scattering between states with different mirror eigenvalues is forbidden under mirror-even scattering potential. As shown in Fig.~\ref{main_fig_normal_QPI} (a) and (c) displays the QPI pattern without the $\gamma$-FS, where intra-FS scattering labeled in Fig.~\ref{main_fig_bilayer_structure}(c) dominates. 
Specifically, the interlayer scattering potential $V_3$, which primarily affects the $d_{z^2}$ orbital, suppresses intra-FS scattering on the $\alpha$ and $\beta$ FSs along the diagonal direction, where these pockets are of pure $d_{x^2-y^2}$-orbital character.
Moreover, although both $V_1$ and $V_3$ are both mirror-even, the values of $\delta \rho(\mathbf{q}, \omega)$ they produce from scattering between antibonding Fermi surfaces are opposite in sign. This difference arises from the distinct ways in which $V_1$ and $V_3$ couple to the antibonding Fermi surfaces, $\langle u_-(\kk)|V_{1/3}|u_-(\kk)\rangle=\pm1$.
The contrast in response manifests in the intensity (absolute value) of the QPI pattern as a cancellation between scattering contributions from the bonding and antibonding FSs when two scattering vectors are close.
For mirror-odd type of scattering potential $V_2$, the QPI intensity $|\delta\rho_2(\qqq,\omega)|$ vanishes throughout the entire Brillouin zone.
The reason is that the numerator of Eq.~\eqref{main_eq_normalQPI} vanishes,
$\Tr(|u_m(\kk')\rangle\langle u_m(\kk')|\ell_z\sigma_0|u_{n}(\kk)\rangle\langle u_{n}(\kk)|)=\delta_{m,-n}\delta_{m,n}$, with $|u_{m/n}(\kk)\rangle$ being the wavefunction with a mirror eigenvalue $m,n=\pm$. 
To address this, we define a ``layer-dependent'' LDOS to quantify the difference between the LDOS on the two layers $\delta\rho^\prime(\qqq,\omega)=-\frac{1}{N\pi}\text{Tr}[\ell_z\sigma_0\sum_{\kk} \delta G(\kk,\kk+\qqq,\omega)]$.
In contrast to $V_{1,3}$, the mirror-odd scattering potential $V_2$ determines that the scattering only occurs between states with opposite mirror eigenvalues. Consequently, the interpocket scattering between $\alpha$ and $\beta$ FSs contributes to peaks around $\qqq_{1,2}$ and all intrapocket scattering is absent, as shown in Fig.~\ref{main_fig_normal_QPI}(b).
When the $\gamma$-FS is present, scattering related $\gamma$-FS is significantly strong due to its large DOS stemming from the flat band near the M point. In Fig.~\ref{main_fig_normal_QPI}(d) and (f), intra-FS scattering within the $\gamma$-FS, namely $\qqq_{2,1}$ in Fig.~\ref{main_fig_normal_QPI}(d),  is dominant and almost masks contributions from $\alpha$ and $\beta$-FSs. For the case of $V_2$ shown in Fig.~\ref{main_fig_normal_QPI}(e), the scattering between $\beta$ and $\gamma$ FSs contributes additional peaks around $\qqq_{2,4}$ and $\qqq_{2,6}$ compared to the case without the $\gamma$-FS.
The selection rules for impurity scattering arising from the mirror symmetry in the normal state are listed in Table~\ref{main_table_scattering_matrix}, which can be analytically derived from the single-band bilayer model (see Supplementary Material).

In Supplementary Material (SM), we present QPI patterns for various renormalized impurity potentials from the $T$-matrix approximation as described in Eq.~\eqref{Tmatrixpotential}. The QPI patterns for the mirror-even potentials $V_{1,3}$ resemble those shown in Fig.~\ref{main_fig_normal_QPI} (a), (d), (c) and (f), while the patterns for the mirror-odd potential $V_2$ display both intra- and inter-pocket scattering.
This behavior arises because if the bare impurity potential preserves a symmetry operation $\hat{L}$ (such as mirror symmetry $\mathcal{M}_z$ in our case), the renormalized potential $T(\omega)$ will also preserve this symmetry, containing only the $\hat{L}$-even component. Conversely, if the bare potential is $\hat{L}$-odd and breaks this symmetry, the renormalized potential $T(\omega)$ will generally include both $\hat{L}$-even and $\hat{L}$-odd components. Consequently, the QPI patterns for the renormalized mirror-even potentials $V_{1,3}$ do not incorporate contributions from $V_2$, whereas those for the renormalized mirror-odd potential $V_2$ include contributions from both $V_2$ and $V_{1,3}$.
The symmetries of the renormalized $T$-matrix are summarized in the last column of Table~\ref{main_table_scattering_matrix}.
Therefore, the QPI patterns in bilayer nickelates exhibit intriguing mirror-selective features, which can be utilized to distinguish different Fermiologies and their associated orbital characters in the normal states.

\textit{QPI diagnosis of pairing symmetries.--}
We further explore the QPI signals in the superconducting states. So far, the pairing symmetry of bilayer nickelates has been under debate~\cite{PhysRevLett.134.136002,PhysRevLett.132.106002,PhysRevB.111.174506,PhysRevB.108.L140505,PhysRevLett.131.236002,10.1038/s41467-024-46622-z,10.1038/s41467-025-56206-0,PhysRevB.108.165141,PhysRevLett.132.036502,PhysRevB.109.165154,PhysRevB.108.L201121,PhysRevB.108.214522,PhysRevLett.132.146002,PhysRevB.108.174511,PhysRevB.108.L201108,PhysRevB.110.024514,PhysRevB.109.104508,PhysRevB.109.L180502,PhysRevB.109.L180502,Jiang_2024,PhysRevLett.132.126503}, and various pairing states, such as interlayer $s_{\pm}$-, intralayer $d_{x^2-y^2}$- and $d_{xy}$-wave pairings, has been theoretically proposed. However, experimental determination of the pairing symmetry remains elusive. QPI provides a powerful method to identify the pairing symmetry in STM measurements, especially due to mirror-selective features in bilayer nickelates. 
We focus on the QPI patterns associated with three representative candidate pairing potentials $\hat{\mathcal{H}}_{\Delta}=\sum_{\kk,\ell\ell^\prime,oo^\prime}\Delta_{\ell\ell^\prime,oo^\prime}(\kk)\hat{c}^\dagger_{\kk,\ell,o}\hat{c}^\dagger_{-\kk,\ell^\prime,o^\prime}+h.c.$: intralayer intraorbital $s$-wave 
\begin{equation}\label{main_eq_swave}
\Delta_{s}(\kk)=\Delta_0\ell_0\sigma_0,
\end{equation}
interlayer intraorbital $s_\pm$-wave 
\begin{equation}
	\Delta_{s\pm}(\kk)=\Delta_0\ell_x\sigma_0,
\end{equation}
and intralayer $d_{x^2-y^2}$-wave pairing
\begin{equation}
	\Delta_{d}(\kk)=2\Delta_0(\cos k_x-\cos k_y)\ell_0\sigma_0.
\end{equation}
The corresponding Bogoliubov-de Gennes (BdG) Hamiltonian reads $\hat{\mathcal{H}}_{\text{BdG}}=\hat{\mathcal{H}}_{0}+\hat{\mathcal{H}}_{\Delta}$. The intralayer $s$-wave pairing results in uniform pairing without sign changes in the band space. In contrast, the interlayer pairing generates sign-reversed gaps on the bonding ($\alpha,\gamma$) and antibonding ($\beta$) FSs, while the intralayer $d$-wave pairing leads to sign-reversed gaps between the $\alpha$- and $\beta$-FSs, in addition to a $d$-wave gap on each FS.
Using this Hamiltonian, we can calculate the even and odd fluctuations of LDOS~\cite{PhysRevB.80.144514}:
\begin{equation}
\left\{
\begin{array}{c}
\delta\rho^{\text{even}}\\
\delta\rho^{\text{odd}}
\end{array}
\right\}
\sim \Im \Tr \sum_{\kk}
\left\{
\begin{array}{c}
\tau_0\\
\tau_z
\end{array}
\right\}
\delta\mathcal{G}(\kk,\kk+\qqq,\omega),
\end{equation}
with $\delta\mathcal{G}(\kk,\kk+\qqq,\omega)=\mathcal{G}_0(\kk,\omega)\mathcal{T}(\omega)\mathcal{G}_0(\kk+\qqq,\omega)$.
Here $\tau_i$ are the Pauli matrices in the Nambu space and $\mathcal{G}$ and $\mathcal{T}$ are the Green function and $T$-matrix in the BdG basis.
To ensure that the coherence factors arising from the quasiparticle wavefunctions and the scattering matrix elements constructively interfere, we should adopt the odd fluctuation for nonmagnetic impurities $\tau_3$ and the even fluctuation for magnetic impurities $\tau_0$ (See SM).
The nonmagnetic impurity is sensitive to sign changes in the superconducting gaps and thus provide crucial information for identifying the pairing symmetry. The full selection rules for impurity scattering in the superconducting state are summarzied in Table~\ref{main_table_scattering_matrix}.
In the following, we focus on the QPI pattern induced by nonmagnetic impurities for two types of Fermiologies, with those from magnetic impurities shown in SM.

Fig.~\ref{main_fig_wog_BdG_QPI} shows the odd fluctuation of QPI intensities induced by nonmagnetic impurities without the $\gamma$-FS.
In the nonmagnetic scattering processes, peaks at $\qqq$-vectors connecting regions of the FS with the same gap sign are suppressed.
Consequently, for the $s$-wave pairing, as shown in Figs.~\ref{main_fig_wog_BdG_QPI} (b)-(d), all peak intensities are weak, and the locations of the remaining peaks closely resemble those observed in the normal state.
In the case of $s_{\pm}$-wave pairing, the nature of interlayer pairing enforces a strict locking between the mirror eigenvalue of each Fermi surface and the pairing sign on it, an effect we refer to as \textit{mirror-pairing locking}.
Thus, mirror-even impurities $V_1$ and $V_3$ only allow intra-FS scattering, leading to QPI patterns that are indistinguishable from the $s$-wave case.
This behavior contrasts significantly with the QPI observed in cuprate and iron-based superconductors~\cite{doi:10.1126/science.1072640,McElroy2003,hanaguri2009coherence,Kohsaka2008,Allan2013,hanaguri2010unconventional}, highlighting the unique features of the interlayer pairing within the bilayer structure.
However, for mirror-odd impurity $V_2$, the allowed scattering vectors are only inter-FS and connect Fermi points with opposite sign of the order parameter, resulting in strong peaks at the $\qqq_{1,2}$ and $\qqq_{1,3}$ in Fig.~\ref{main_fig_wog_BdG_QPI} (g).
Note that in this case, we use the layer-dependent LDOS $\delta \rho^{\text{odd}}\sim \Im\Tr\sum_{\kk}(\tau_z\ell_z\sigma_0)\delta \mathcal{G}_{\kk,\kk+\qqq}$.
For the $d_{x^2-y^2}$-wave pairing, the intra-FS scattering between Fermi points with sign-changed order parameters contributes to peaks under the mirror-even potential $V_{1,3}$, as shown in Figs.~\ref{main_fig_wog_BdG_QPI}(j) and (l). The peak intensity from $V_3$ is suppressed due to its dominant $d_{z^2}$-orbital scattering.  With the mirror-odd potential $V_2$, only inter-FS scattering can contribute and the resulting QPI is shown in Fig.~\ref{main_fig_wog_BdG_QPI}(k). It is similar to the mirror-even potential due to their close scattering vectors.

\begin{figure}[t]
	\centering
	\includegraphics[width=0.9875\linewidth]{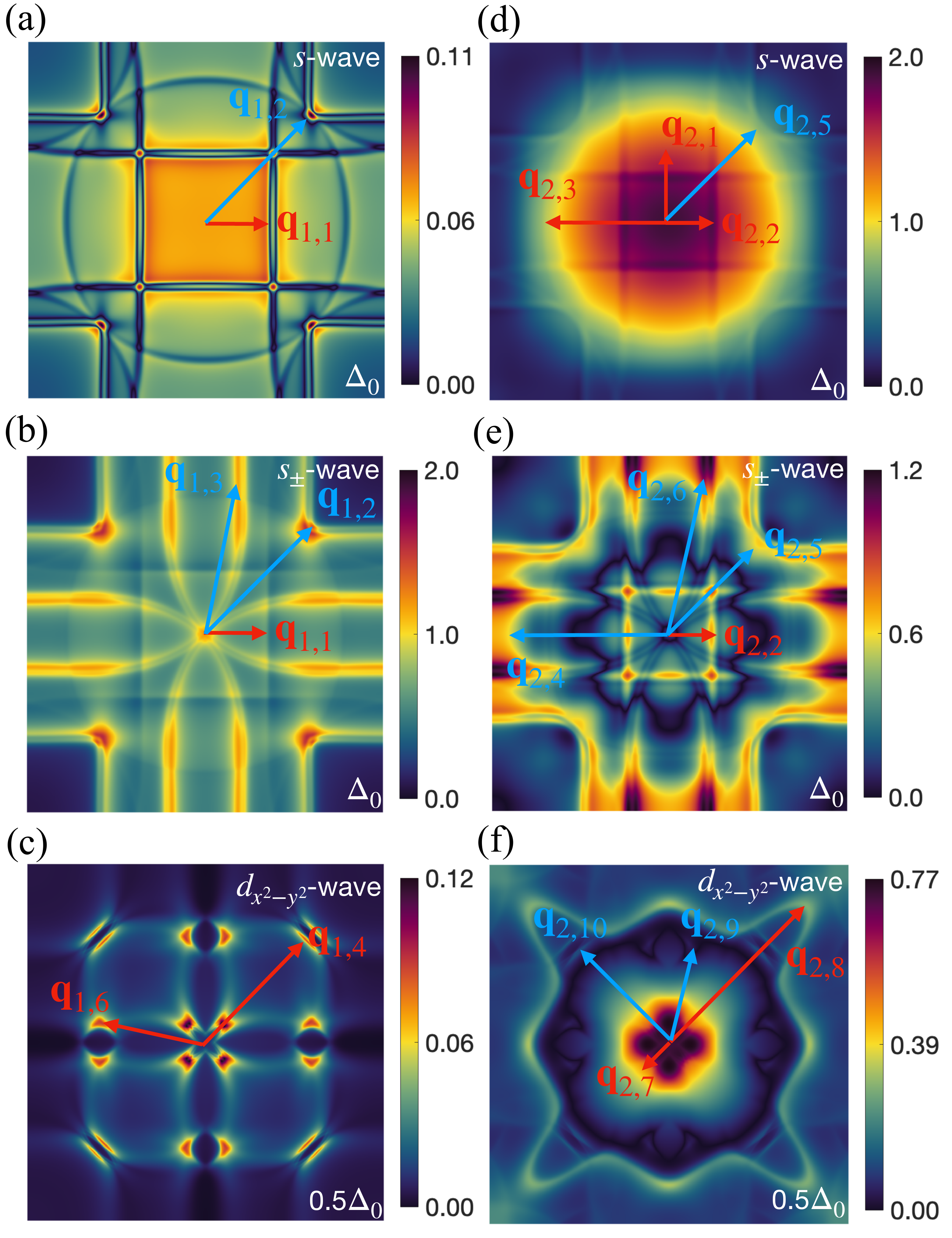}
	\caption{\label{main_fig_BdG_toplayer_QPI} {\bf The QPI intensity $Z(\qqq,\omega)$ on the top layer with renormalized impurities potential $T_t(\omega)$.}
	(a)-(c) $Z(\qqq,\omega)$ for three representative pairing symmetries without the $\gamma$-FS.
	For $s$ and $s_\pm$-wave pairing, the QPI are measured at $\omega=\Delta_0$.
	For $d_{x^2-y^2}$-wave pairing, the QPI are measured at $\omega=\Delta_0/2$.
	(d)-(f) the same quantities but for the case with the $\gamma$-FS.
	}
\end{figure}

In the presence of the $\gamma$-FS, the QPI patterns for three representative pairing states are presented in the SM.
Similar to the normal state, the scattering related to the $\gamma$-FS dominates, contributing higher intensity. Apart from this, the main features for both $s$- and $d_{x^2-y^2}$-wave pairings resemble those in Fig.~\ref{main_fig_wog_BdG_QPI}. Notably, for the $d_{x^2-y^2}$-wave pairing, inter-pocket scattering involving the $\gamma$-FS is weak at $\omega=\Delta_0/2$ due to its small superconducting gap. While, the $s_{\pm}$-wave pairing shows an additional peak at $\qqq_{2,4}$ in the $V_2$ potential scattering, which is attributed to scattering between the $\beta$ and $\gamma$ FSs.

We further consider the situation in practical STM measurements. Since STM is a local probe, the measured LDOS primarily reflects that of the top layer. We therefore consider two natural cases: impurities resides either between the bilayers or on the top layer. In first case, the impurity potential is just $V_3$ and the QPI pattern on the top layer is shown in Figs.~\ref{main_fig_wog_BdG_QPI}(d), (h) and (l). In the latter case, the potential corresponds to the combination of mirror-even $V_1$
and mirror-odd $V_2$. The FT-QPI pattern on the top layer reads $Z(\qqq,\omega) \sim\Im\Tr\sum_{\kk}\tau_3\ell_t\delta\mathcal{G}_{\kk,\kk+\qqq}
$ with $T_t(\omega)=\tau_3\ell_t(1-\frac{1}{N}\sum_{\kk}\mathcal{G}_{0}(\kk,\omega)\tau_3\ell_t)^{-1}$ and $\ell_t=(\ell_0+\ell_3)/2$.
Fig.~\ref{main_fig_BdG_toplayer_QPI} shows the corresponding LDOS in momentum space for three pairing states and two Fermiologies with a nonmagnetic impurity located in the top layer.
As expected, both intra-FS and inter-FS scatterings appear in the QPI patterns. It is apparent that different pairings exhibit distinct QPI patterns, enabling the differentiation of pairing states through STM measurements.

\begin{figure}[b]
	\centering
	\includegraphics[width=0.9875\linewidth]{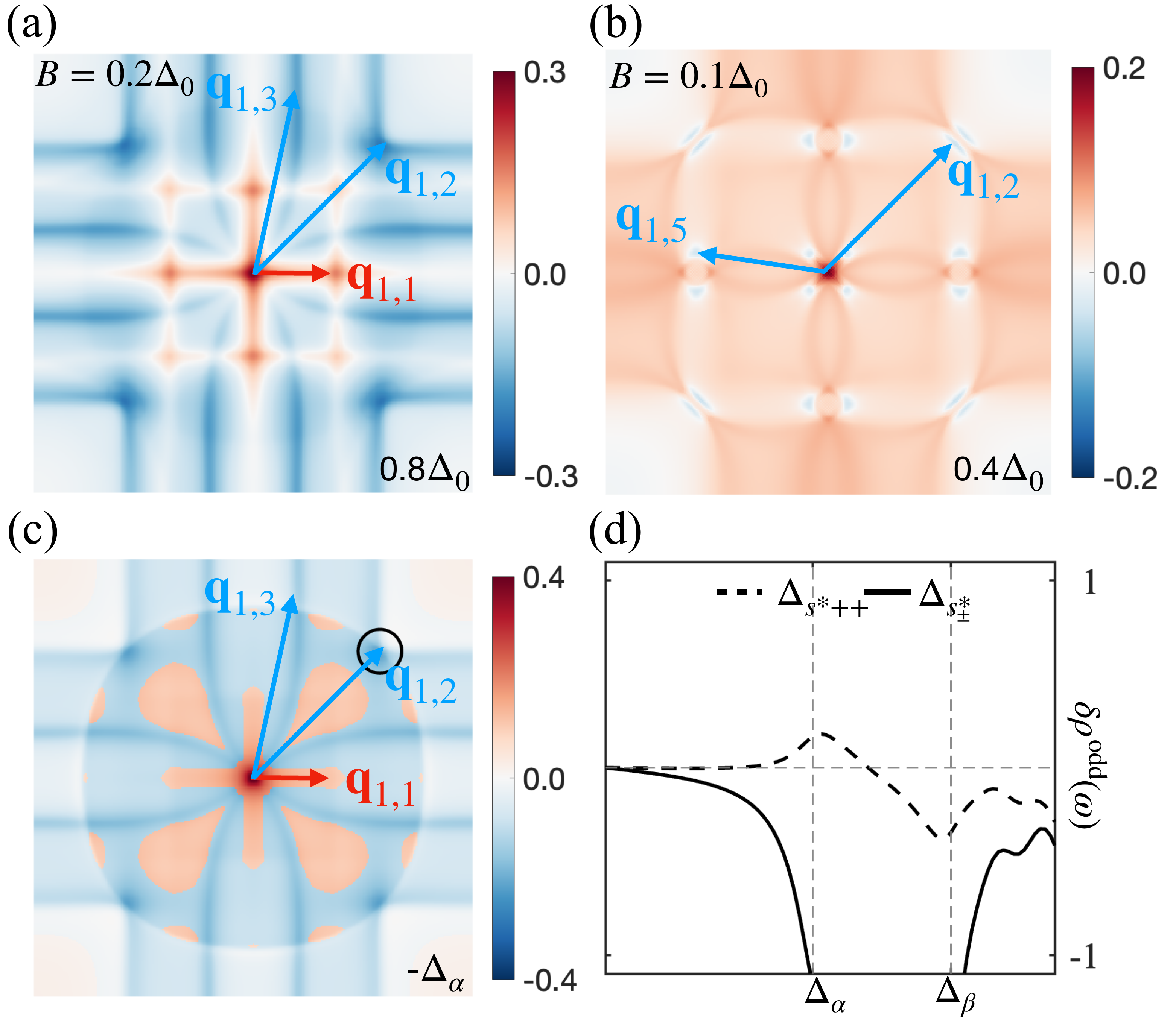}
	\caption{\label{main_fig_BdG_toplayer_FieldQPIandRQPI} {\bf Field dependent and reference QPI} (a)-(b) Magnetic field-induced change in QPI pattern $|Z_{B=B_0}(\qqq,\omega)|-|Z_{B=0}(\qqq,\omega)|$ for $s_\pm$-wave and $d_{x^2-y^2}$-wave pairing.
	The strength of the nonmagnetic impurity is $v_0=0.25$, and the strength ratio between the vortex impurity and the nonmagnetic impurity is $2$.
	(c) The PR-QPI intensity $\delta \rho_{\text{PR}}(\qqq,\omega=-|\Delta_\alpha|)$ induced by top layer nonmagnetic impurity in Eq.~\eqref{main_eq_PRQPI} for $s_\pm^*$-wave pairing without hole doping.
(d) The integrated QPI intensity $\delta\rho^{\text{odd}}$ around $\qqq_{1,2}$ for for $s^*_{++}$-wave and $s_\pm^*$-wave pairing.
The parameters are set to be $\{\Delta_\alpha,\Delta_\beta\}=\{0.6\Delta_0,\Delta_0\}$.
	}
\end{figure}

\textit{Field dependent and reference QPI.--}
In STM measurements, factors such as nature of impurities, multi-orbital induced FS dependent gaps, and gap anisotropy in bilayer nickelates can affect QPI signals and obscure selection rules discussed earlier~\cite{PhysRevB.92.045124}. To accurately identify the sign changes in the superconducting gaps,  we examine the magnetic field dependence of QPI patterns~\cite{hanaguri2010unconventional,PhysRevB.67.020511,PhysRevB.84.054501} and employ phase-references QPI techniques~\cite{PhysRevB.92.184513}.
Applying a magnetic field introduces Zeeman splitting in the electronic structure and generate vortices, acting as additional scatters.
An out-of-plane magnetic field maintains mirror symmetry $\mathcal{M}_z$ but breaks time-reversal symmetry. We consider the following process to simulate the magnetic field dependent QPI: at zero magnetic field $B$, impurities contribute nonmagnetic scattering.
As the magnetic field becomes finite, the generated vortices become pinned at the impurity sites and mask impurity scattering, contributing additional scattering in the $\tau_{1}$ Nambu channel.
Note that for the scattering processes in particle-particle channel, the fluctuation of LDOS in the even channel is dominant in $|Z_{B=B_0}(\qqq,\omega)|$~\cite{PhysRevB.78.020509,PhysRevB.80.144514}.
In Figs.~\ref{main_fig_BdG_toplayer_FieldQPIandRQPI} (a) and (b), we show the magnetic field-induced change of QPI $|Z_{B=B_0}(\qqq,\omega)|-|Z_{B=0}(\qqq,\omega)|$.
For both $s_{\pm}$-wave and $d_{x^2-y^2}$-wave pairing without the $\gamma$-FS, the magnetic field suppresses the peaks at $\mathbf{q}$-vectors connecting the order parameter with opposite signs, such as $\qqq_{1,2/3}$ and $\qqq_{1,2/5}$, but slightly enhance those connecting the order parameter with same sign, like $\qqq_{1,1}$. Similar behaviors also appear for the case with the $\gamma$-FS (see SM). The distinct magnetic field responses of $\mathbf{q}$-vectors connecting regions with the identical versus opposite signs of the order parameter can provide a useful criterion for experimental identification of pairing states~\cite{hanaguri2009coherence}.

When the gap amplitudes of two FSs connected by a scattering vector are unequal, determining their phase difference using conventional QPI methods becomes challenging. 
Instead, we can use phase-reference QPI (PR-QPI) to distinguish $s_{++}$- and $s_{\pm}$-wave pairings~\cite{chi2017extracting,chi2017determination}, in analogous to iron based superconductors. With including both intralayer and interlayer pairing, we introduce two pairings with different gap amplitudes on bonding and antibonding FSs: $\Delta_{s^*_{++}}=\frac{1}{2}(|\Delta_\alpha|+|\Delta_\beta|)\ell_0\sigma_0+\frac{1}{2}(|\Delta_\alpha|-|\Delta_\beta|)\ell_x\sigma_0$ and $\Delta_{s^*_\pm}=\frac{1}{2}(|\Delta_\alpha|+|\Delta_\beta|)\ell_x\sigma_0+\frac{1}{2}(|\Delta_\alpha|-|\Delta_\beta|)\ell_0\sigma_0$.
The corresponding density of states are shown in the Supplementary Material.
PR-QPI involves analyzing QPI data by considering both the amplitude and phase of the the LDOS modulations.
In general, the quantity $\delta\rho(\qqq,\omega)$ obtained from Fourier transformation of the real-space LDOS is complex.
By comparing the relative phases of the QPI data at positive and negative energies, one can extract information about the pairing symmetry:
\begin{equation}\label{main_eq_PRQPI}
\begin{split}
	    \delta \rho_{\text{PR}}(\qqq,\omega>0)&\equiv|\delta\rho(\qqq,\omega)|,\\
    \delta \rho_{\text{PR}}(\qqq,-\omega<0)&\equiv|\delta\rho(\qqq,-\omega)|\Re[e^{i(\theta_{\qqq,-\omega}-\theta_{\qqq,\omega})}],
\end{split}
\end{equation}
where $\theta_{\qqq,\omega}$ is the phase of $\delta\rho(\qqq,\omega)\sim\Im\Tr\sum_{\kk}(\tau_0+\tau_3)\ell_t\delta\mathcal{G}_{\kk,\kk+\qqq}
$ with $T_t(\omega)$.
We plot $\delta \rho_{\text{PR}}(\qqq,-\Delta_\alpha)$ in Fig.~\ref{main_fig_BdG_toplayer_FieldQPIandRQPI}(c) for the $s_{\pm}^*$-wave case. The negative values at $\qqq_{1,2/3}$ for the inter-FS scattering suggest that the QPI signals at $+E$ and $-E$ exhibit a sign reversal, consistent with sign-changing gaps on $\alpha$ and $\beta$ FSs. The sign-changed gap can become transparent upon integrating $\delta\rho$ around these $\qqq$ vectors as a function of energy,
\begin{equation}
	\delta\rho^{\text{odd}}(\omega)=\int_{\qqq\in \bigcirc}\dd\qqq\ \Im\Tr\sum_{\kk} \tau_z\ell_t\delta \mathcal{G}_{\kk,\kk+\qqq}.
\end{equation}
Fig.~\ref{main_fig_BdG_toplayer_FieldQPIandRQPI}(d) shows $\delta\rho^\text{odd}(\omega)$ for both $s^*_{++}$- and $s^*_{\pm}$-wave pairings. For the $s^*_{++}$ case, with increasing $\omega$, the integrated value undergoes a sign change and crosses zero at the energy corresponding to the average of the two gap magnitudes. In contrast, $\delta\rho^\text{odd}(\omega)$ exhibit no sign change between $\Delta_\alpha$ and $\Delta_\beta$. These distinct behaviors provide a clear method to distinguish between $s^*_{++}$ and $s^*_{\pm}$ pairing with FS-dependent gap amplitudes~\cite{PhysRevB.92.184513}.

\textit{Discussion.--}
 The inherent mirror reflection symmetry within bilayer nickelates allows for the mirror-selective detection of FSs and superconducting gaps through STM measurements. The distinct QPI patterns observed with and without the $\gamma$-FS can help resolve debates over the Fermiology of bilayer nickelate thin films in the normal state. In the superconducting state, different types of impurity scattering can selectively identify the sign changes of gaps within and between FSs. Even in the presence of gap anisotropy, these symmetry-related scattering features, along with field-dependent and reference QPI, enable robust determination of pairing symmetry.
Due to the multiorbital nature, interlayer repulsion can induce an interlayer interorbital $d$-wave pairing $\Delta_0=\ell_x\sigma_x$, which differs from the previously mentioned intralayer $d$-wave pairing. These two $d$-wave pairings display similar sign changes of gaps across different FSs, making them difficult to distinguish. When the $\gamma$ FS is present, the interlayer interorbital pairing is almost vanishing on this $d_{z^2}$-dominant FS, whereas the intralayer $d$-wave pairing remains finite. This distinction could be resolved via ARPES or transport measurements.

For practical STM measurements, it is essential to target different regions to access different types of impurities. Accurate determination of pairing symmetry requires observing both mirror-even and mirror-odd impurity scatterings; otherwise, incorrect conclusions might be drawn. For instance, considering only the mirror-even scattering $V_{1,3}$ for the interlayer $s_{\pm}$ pairing, as shown in Fig.~\ref{main_fig_wog_BdG_QPI}, both intra- and inter-FS scattering are absent in QPI patterns, which might incorrectly suggest an $s$-wave superconducting gap with no sign change.
However, this absence is due to the mirror-pairing locking, which arises from the unique interlayer pairing.
As a result, gap sign changes between FSs can only be inferred from mirror-odd scattering $V_2$, clearly distinguishing between $s$-wave and $s_{\pm}$-wave pairings.
Note that such features do not appear in the intralayer $s_{\pm}$-pairing, as seen in iron-based superconductors, where the mirror eigenvalue and the paring sign are decoupled.
When impurities are present on either the top or bottom layer, both intra- and inter-FS scattering can be detected in the QPI, similar to the cases in cuprates and iron-based superconductors, allowing for conventional QPI interpretation. 
In addition, if both types of mirror-even impurities, i.e. $V_1$ and $V_3$, are present in the sample simultaneously, their contributions for QPI can cancel with each other out due to their opposite responses to the antibonding $\beta$-FS. 
As a result, the QPI signal from the antibonding $\beta$-FS becomes nearly silent.
Our analysis of QPI patterns also extends to trilayer nickelates in the high-symmetry phase, where both mirror-even and mirror-odd $\beta$ pockets appear. The distinctive bilayer structure of nickelate superconductors, combined with their nontrivial mirror symmetry, offers a powerful approach to identify the pairing symmetry through QPI.

In summary, we demonstrate that the inherent mirror symmetry of the bilayer nickelate dictates mirror-selective quasiparticle scattering, governed by selection rules related to the mirror nature of impurities and the mirror eigenvalues of electronic wavefunctions. This selectivity provides a powerful dual capability: it enables differentiation two different Fermiologies and allows for separate detection of superconducting gap sign changes within the same FS and between different FSs. When combined with field-dependent and reference QPI measurements, this approach facilitates the precise determination of pairing symmetry, robust against complications from FS-dependent gaps and gap anisotropy.
Our findings establish mirror-selective QPI as an essential tool for resolving the debated Fermiology and distinguishing between proposed pairing states in bilayer nickelates, providing critical insights for pinning down the pairing symmetry and ultimately revealing the superconducting pairing mechanism.

\textit{Funding Declaration.--}
Z.Z. and H.C.P. acknowledge support from the Croucher Foundation through CIA23SC01 and the Hong Kong Research Grants Council through ECS 26308021. J.H. acknowledge the supports by the Ministry of Science and Technology (Grant No. 2022YFA1403901), National Natural Science Foundation of China (No. 11920101005, No. 11888101, No. 12047503, No. 12322405, No. 12104450) and the New Cornerstone Investigator Program.
X.W. is supported by the National Key R$\&$D Program of China (Grant No. 2023YFA1407300) and the National Natural Science Foundation of China (Grants No. 12447103).

\textit{Data availability} All data needed to evaluate the conclusions in the study are present in
the paper and/or the Supplementary Information. The data that support the findings of this study are available from the corresponding authors upon request.

\textit{Code availability} The computer code used for numerical calculation and theoretical understanding is available upon request from the corresponding authors.

\bibliography{ref}

\textit{Author contributions:} Z.Z., J.Z., H.C.P. and X.W. did the theoretical derivation and numerical calculation; C.L. did DFT calculation. Z.Z., H.C.P., J.H. and X.W. provided the theoretical understanding. All authors discussed and contributed to the manuscript. Z.Z, J.H. and X.W. conceived the work.

\textit{Competing interests:} The authors declare no competing interests.

\clearpage

\end{document}

%% file: preamble.tex
\usepackage{epsf}
\usepackage{here,times}
\usepackage[colorlinks=true,citecolor=blue,linkcolor=blue, urlcolor=blue]{hyperref} 
\usepackage{slashed}
\usepackage{amsthm}
\usepackage{thmtools}
\usepackage{blindtext}
\usepackage{physics, bm}
\usepackage{mathrsfs}
\usepackage{amsfonts, mathtools, amsmath,amssymb,dsfont}
\usepackage{esvect}
\usepackage{enumerate,dcolumn,multirow,bbold, enumitem}
\usepackage{longtable}
\usepackage{array}
\usepackage{tikz-cd}
\usepackage{bbding}
\usepackage[normalem]{ulem}
\usepackage{mathrsfs}
\usepackage{calligra}
\usepackage{multirow}
\usepackage{bbm}
\usepackage{mathbbol}

\usepackage{algorithm, algorithmic}

\declaretheoremstyle[
shaded={bgcolor=\color{rgb}{0.9,0.9,0.9}}  
]{theorem}

\declaretheoremstyle[
shaded={bgcolor=\color{rgb}{0.9,0.9,0.9}}
]{question}

\declaretheoremstyle[
shaded={bgcolor=\color{rgb}{0.9,0.9,0.9}}  
]{remark}

\declaretheoremstyle[
shaded={bgcolor=\color{rgb}{0.9,0.9,0.9}}  
]{proposition}

\declaretheoremstyle[
shaded={bgcolor=\color{rgb}{0.9,0.9,0.9}}  
]{definition}

\declaretheoremstyle[
shaded={bgcolor=\color{rgb}{0.9,0.9,0.9}}  
]{assumption}

\declaretheoremstyle[
shaded={bgcolor=\color{rgb}{0.9,0.9,0.9}}  
]{conjecture}

\declaretheoremstyle[
shaded={bgcolor=\color{rgb}{0.9,0.9,0.9}}  
]{corrorary}

\declaretheoremstyle[
shaded={bgcolor=\color{rgb}{0.9,0.9,0.9}}  
]{axiom}

\declaretheoremstyle[
shaded={bgcolor=\color{rgb}{0.9,0.9,0.9}}  
]{lemma}

\declaretheoremstyle[
shaded={bgcolor=\color{rgb}{0.9,0.9,0.9}}  
]{problem}

\usepackage{epsf}
\usepackage{here,times}
\usepackage{slashed}
\usepackage{amsthm}
\usepackage{thmtools}
\usepackage{blindtext}
\usepackage{physics, bm}
\usepackage{mathrsfs}
\usepackage{amsfonts, mathtools, amsmath,amssymb,dsfont}
\usepackage{esvect}
\usepackage{enumerate,dcolumn,multirow,bbold, enumitem}
\usepackage{longtable}
\usepackage{array}
\usepackage{tikz-cd}
\usepackage{bbding}
\usepackage[normalem]{ulem}
\usepackage{mathrsfs}
\usepackage{calligra}

\newcommand{\RNum}[1]{\uppercase\expandafter{\romannumeral #1\relax}}

\newcommand{\Rom}[1]{ \uppercase\expandafter{\romannumeral#1}}

\newcommand{\kk}{{\bf{{k}}}}
\newcommand{\qqq}{{\bf{{q}}}}

\newcommand{\rr}{{\bf{r}}}

\definecolor{ZZYcolor}{rgb}{0.1,0.5,0.4}

%% file: main.bbl
\begin{thebibliography}{81}%
\makeatletter
\providecommand \@ifxundefined [1]{%
 \@ifx{#1\undefined}
}%
\providecommand \@ifnum [1]{%
 \ifnum #1\expandafter \@firstoftwo
 \else \expandafter \@secondoftwo
 \fi
}%
\providecommand \@ifx [1]{%
 \ifx #1\expandafter \@firstoftwo
 \else \expandafter \@secondoftwo
 \fi
}%
\providecommand \natexlab [1]{#1}%
\providecommand \enquote  [1]{``#1''}%
\providecommand \bibnamefont  [1]{#1}%
\providecommand \bibfnamefont [1]{#1}%
\providecommand \citenamefont [1]{#1}%
\providecommand \href@noop [0]{\@secondoftwo}%
\providecommand \href [0]{\begingroup \@sanitize@url \@href}%
\providecommand \@href[1]{\@@startlink{#1}\@@href}%
\providecommand \@@href[1]{\endgroup#1\@@endlink}%
\providecommand \@sanitize@url [0]{\catcode `\\12\catcode `\$12\catcode `\&12\catcode `\#12\catcode `\^12\catcode `\_12\catcode `\%12\relax}%
\providecommand \@@startlink[1]{}%
\providecommand \@@endlink[0]{}%
\providecommand \url  [0]{\begingroup\@sanitize@url \@url }%
\providecommand \@url [1]{\endgroup\@href {#1}{\urlprefix }}%
\providecommand \urlprefix  [0]{URL }%
\providecommand \Eprint [0]{\href }%
\providecommand \doibase [0]{https://doi.org/}%
\providecommand \selectlanguage [0]{\@gobble}%
\providecommand \bibinfo  [0]{\@secondoftwo}%
\providecommand \bibfield  [0]{\@secondoftwo}%
\providecommand \translation [1]{[#1]}%
\providecommand \BibitemOpen [0]{}%
\providecommand \bibitemStop [0]{}%
\providecommand \bibitemNoStop [0]{.\EOS\space}%
\providecommand \EOS [0]{\spacefactor3000\relax}%
\providecommand \BibitemShut  [1]{\csname bibitem#1\endcsname}%
\let\auto@bib@innerbib\@empty
\bibitem [{\citenamefont {Sun}\ \emph {et~al.}(2023)\citenamefont {Sun}, \citenamefont {Huo}, \citenamefont {Hu}, \citenamefont {Li}, \citenamefont {Liu}, \citenamefont {Han}, \citenamefont {Tang}, \citenamefont {Mao}, \citenamefont {Yang}, \citenamefont {Wang}, \citenamefont {Cheng}, \citenamefont {Yao}, \citenamefont {Zhang},\ and\ \citenamefont {Wang}}]{sun2023}%
  \BibitemOpen
  \bibfield  {author} {\bibinfo {author} {\bibfnamefont {H.}~\bibnamefont {Sun}}, \bibinfo {author} {\bibfnamefont {M.}~\bibnamefont {Huo}}, \bibinfo {author} {\bibfnamefont {X.}~\bibnamefont {Hu}}, \bibinfo {author} {\bibfnamefont {J.}~\bibnamefont {Li}}, \bibinfo {author} {\bibfnamefont {Z.}~\bibnamefont {Liu}}, \bibinfo {author} {\bibfnamefont {Y.}~\bibnamefont {Han}}, \bibinfo {author} {\bibfnamefont {L.}~\bibnamefont {Tang}}, \bibinfo {author} {\bibfnamefont {Z.}~\bibnamefont {Mao}}, \bibinfo {author} {\bibfnamefont {P.}~\bibnamefont {Yang}}, \bibinfo {author} {\bibfnamefont {B.}~\bibnamefont {Wang}}, \bibinfo {author} {\bibfnamefont {J.}~\bibnamefont {Cheng}}, \bibinfo {author} {\bibfnamefont {D.-X.}\ \bibnamefont {Yao}}, \bibinfo {author} {\bibfnamefont {G.-M.}\ \bibnamefont {Zhang}},\ and\ \bibinfo {author} {\bibfnamefont {M.}~\bibnamefont {Wang}},\ }\bibfield  {title} {\bibinfo {title} {{Signatures of superconductivity near 80 K in a nickelate under high pressure}},\ }\href {https://doi.org/10.1038/s41586-023-06408-7} {\bibfield  {journal} {\bibinfo  {journal} {Nature}\ }\textbf {\bibinfo {volume} {621}},\ \bibinfo {pages} {493} (\bibinfo {year} {2023})}\BibitemShut {NoStop}%
\bibitem [{\citenamefont {Luo}\ \emph {et~al.}(2023)\citenamefont {Luo}, \citenamefont {Hu}, \citenamefont {Wang}, \citenamefont {W\'u},\ and\ \citenamefont {Yao}}]{YaoDX}%
  \BibitemOpen
  \bibfield  {author} {\bibinfo {author} {\bibfnamefont {Z.}~\bibnamefont {Luo}}, \bibinfo {author} {\bibfnamefont {X.}~\bibnamefont {Hu}}, \bibinfo {author} {\bibfnamefont {M.}~\bibnamefont {Wang}}, \bibinfo {author} {\bibfnamefont {W.}~\bibnamefont {W\'u}},\ and\ \bibinfo {author} {\bibfnamefont {D.-X.}\ \bibnamefont {Yao}},\ }\bibfield  {title} {\bibinfo {title} {{Bilayer Two-Orbital Model of $\mathrm{L}{\mathrm{a}}_{3}\mathrm{N}{\mathrm{i}}_{2}$O$_{7}$ under Pressure}},\ }\href {https://doi.org/10.1103/PhysRevLett.131.126001} {\bibfield  {journal} {\bibinfo  {journal} {Phys. Rev. Lett.}\ }\textbf {\bibinfo {volume} {131}},\ \bibinfo {pages} {126001} (\bibinfo {year} {2023})}\BibitemShut {NoStop}%
\bibitem [{\citenamefont {Zhang}\ \emph {et~al.}(2023{\natexlab{a}})\citenamefont {Zhang}, \citenamefont {Lin}, \citenamefont {Moreo},\ and\ \citenamefont {Dagotto}}]{YZhang2023}%
  \BibitemOpen
  \bibfield  {author} {\bibinfo {author} {\bibfnamefont {Y.}~\bibnamefont {Zhang}}, \bibinfo {author} {\bibfnamefont {L.-F.}\ \bibnamefont {Lin}}, \bibinfo {author} {\bibfnamefont {A.}~\bibnamefont {Moreo}},\ and\ \bibinfo {author} {\bibfnamefont {E.}~\bibnamefont {Dagotto}},\ }\bibfield  {title} {\bibinfo {title} {{Electronic structure, dimer physics, orbital-selective behavior, and magnetic tendencies in the bilayer nickelate superconductor ${\mathrm{La}}_{3}{\mathrm{Ni}}_{2}{\mathrm{O}}_{7}$ under pressure}},\ }\href {https://doi.org/10.1103/PhysRevB.108.L180510} {\bibfield  {journal} {\bibinfo  {journal} {Phys. Rev. B}\ }\textbf {\bibinfo {volume} {108}},\ \bibinfo {pages} {L180510} (\bibinfo {year} {2023}{\natexlab{a}})}\BibitemShut {NoStop}%
\bibitem [{\citenamefont {Lechermann}\ \emph {et~al.}(2023{\natexlab{a}})\citenamefont {Lechermann}, \citenamefont {Gondolf}, \citenamefont {B\"otzel},\ and\ \citenamefont {Eremin}}]{Lechermann2023}%
  \BibitemOpen
  \bibfield  {author} {\bibinfo {author} {\bibfnamefont {F.}~\bibnamefont {Lechermann}}, \bibinfo {author} {\bibfnamefont {J.}~\bibnamefont {Gondolf}}, \bibinfo {author} {\bibfnamefont {S.}~\bibnamefont {B\"otzel}},\ and\ \bibinfo {author} {\bibfnamefont {I.~M.}\ \bibnamefont {Eremin}},\ }\bibfield  {title} {\bibinfo {title} {{Electronic correlations and superconducting instability in ${\mathrm{La}}_{3}{\mathrm{Ni}}_{2}{\mathrm{O}}_{7}$ under high pressure}},\ }\href {https://doi.org/10.1103/PhysRevB.108.L201121} {\bibfield  {journal} {\bibinfo  {journal} {Phys. Rev. B}\ }\textbf {\bibinfo {volume} {108}},\ \bibinfo {pages} {L201121} (\bibinfo {year} {2023}{\natexlab{a}})}\BibitemShut {NoStop}%
\bibitem [{\citenamefont {Sakakibara}\ \emph {et~al.}(2024{\natexlab{a}})\citenamefont {Sakakibara}, \citenamefont {Kitamine}, \citenamefont {Ochi},\ and\ \citenamefont {Kuroki}}]{Hirofumi2023possible}%
  \BibitemOpen
  \bibfield  {author} {\bibinfo {author} {\bibfnamefont {H.}~\bibnamefont {Sakakibara}}, \bibinfo {author} {\bibfnamefont {N.}~\bibnamefont {Kitamine}}, \bibinfo {author} {\bibfnamefont {M.}~\bibnamefont {Ochi}},\ and\ \bibinfo {author} {\bibfnamefont {K.}~\bibnamefont {Kuroki}},\ }\bibfield  {title} {\bibinfo {title} {{Possible High ${T}_{c}$ Superconductivity in ${\mathrm{La}}_{3}{\mathrm{Ni}}_{2}$O$_{7}$ under High Pressure through Manifestation of a Nearly Half-Filled Bilayer Hubbard Model}},\ }\href {https://doi.org/10.1103/PhysRevLett.132.106002} {\bibfield  {journal} {\bibinfo  {journal} {Phys. Rev. Lett.}\ }\textbf {\bibinfo {volume} {132}},\ \bibinfo {pages} {106002} (\bibinfo {year} {2024}{\natexlab{a}})}\BibitemShut {NoStop}%
\bibitem [{\citenamefont {Gu}\ \emph {et~al.}(2025{\natexlab{a}})\citenamefont {Gu}, \citenamefont {Le}, \citenamefont {Yang}, \citenamefont {Wu},\ and\ \citenamefont {Hu}}]{XWu}%
  \BibitemOpen
  \bibfield  {author} {\bibinfo {author} {\bibfnamefont {Y.}~\bibnamefont {Gu}}, \bibinfo {author} {\bibfnamefont {C.}~\bibnamefont {Le}}, \bibinfo {author} {\bibfnamefont {Z.}~\bibnamefont {Yang}}, \bibinfo {author} {\bibfnamefont {X.}~\bibnamefont {Wu}},\ and\ \bibinfo {author} {\bibfnamefont {J.}~\bibnamefont {Hu}},\ }\bibfield  {title} {\bibinfo {title} {{Effective model and pairing tendency in the bilayer Ni-based superconductor ${\mathrm{La}}_{3}{\mathrm{Ni}}_{2}{\mathrm{O}}_{7}$}},\ }\href {https://doi.org/10.1103/PhysRevB.111.174506} {\bibfield  {journal} {\bibinfo  {journal} {Phys. Rev. B}\ }\textbf {\bibinfo {volume} {111}},\ \bibinfo {pages} {174506} (\bibinfo {year} {2025}{\natexlab{a}})}\BibitemShut {NoStop}%
\bibitem [{\citenamefont {Yang}\ \emph {et~al.}(2024)\citenamefont {Yang}, \citenamefont {Sun}, \citenamefont {Hu}, \citenamefont {Xie}, \citenamefont {Miao}, \citenamefont {Luo}, \citenamefont {Chen}, \citenamefont {Liang}, \citenamefont {Zhu}, \citenamefont {Qu}, \citenamefont {Chen}, \citenamefont {Huo}, \citenamefont {Huang}, \citenamefont {Zhang}, \citenamefont {Zhang}, \citenamefont {Yang}, \citenamefont {Wang}, \citenamefont {Peng}, \citenamefont {Mao}, \citenamefont {Liu}, \citenamefont {Xu}, \citenamefont {Qian}, \citenamefont {Yao}, \citenamefont {Wang}, \citenamefont {Zhao},\ and\ \citenamefont {Zhou}}]{XJZhou2023}%
  \BibitemOpen
  \bibfield  {author} {\bibinfo {author} {\bibfnamefont {J.}~\bibnamefont {Yang}}, \bibinfo {author} {\bibfnamefont {H.}~\bibnamefont {Sun}}, \bibinfo {author} {\bibfnamefont {X.}~\bibnamefont {Hu}}, \bibinfo {author} {\bibfnamefont {Y.}~\bibnamefont {Xie}}, \bibinfo {author} {\bibfnamefont {T.}~\bibnamefont {Miao}}, \bibinfo {author} {\bibfnamefont {H.}~\bibnamefont {Luo}}, \bibinfo {author} {\bibfnamefont {H.}~\bibnamefont {Chen}}, \bibinfo {author} {\bibfnamefont {B.}~\bibnamefont {Liang}}, \bibinfo {author} {\bibfnamefont {W.}~\bibnamefont {Zhu}}, \bibinfo {author} {\bibfnamefont {G.}~\bibnamefont {Qu}}, \bibinfo {author} {\bibfnamefont {C.-Q.}\ \bibnamefont {Chen}}, \bibinfo {author} {\bibfnamefont {M.}~\bibnamefont {Huo}}, \bibinfo {author} {\bibfnamefont {Y.}~\bibnamefont {Huang}}, \bibinfo {author} {\bibfnamefont {S.}~\bibnamefont {Zhang}}, \bibinfo {author} {\bibfnamefont {F.}~\bibnamefont {Zhang}}, \bibinfo {author} {\bibfnamefont {F.}~\bibnamefont {Yang}}, \bibinfo {author} {\bibfnamefont {Z.}~\bibnamefont {Wang}}, \bibinfo {author} {\bibfnamefont {Q.}~\bibnamefont {Peng}}, \bibinfo {author} {\bibfnamefont {H.}~\bibnamefont {Mao}}, \bibinfo {author} {\bibfnamefont {G.}~\bibnamefont {Liu}}, \bibinfo {author} {\bibfnamefont {Z.}~\bibnamefont {Xu}}, \bibinfo {author} {\bibfnamefont {T.}~\bibnamefont {Qian}}, \bibinfo {author} {\bibfnamefont {D.-X.}\ \bibnamefont {Yao}}, \bibinfo {author} {\bibfnamefont {M.}~\bibnamefont {Wang}}, \bibinfo {author} {\bibfnamefont {L.}~\bibnamefont {Zhao}},\ and\ \bibinfo {author} {\bibfnamefont {X.~J.}\ \bibnamefont {Zhou}},\ }\bibfield  {title} {\bibinfo {title} {{Orbital-dependent electron correlation in double-layer nickelate {La$_3$Ni$_2$O$_7$}}},\ }\href {https://doi.org/10.1038/s41467-024-48701-7} {\bibfield  {journal} {\bibinfo  {journal} {Nature Communications}\ }\textbf {\bibinfo {volume} {15}},\ \bibinfo {pages} {4373} (\bibinfo {year} {2024})}\BibitemShut {NoStop}%
\bibitem [{\citenamefont {Liu}\ \emph {et~al.}(2024)\citenamefont {Liu}, \citenamefont {Huo}, \citenamefont {Li}, \citenamefont {Li}, \citenamefont {Liu}, \citenamefont {Dai}, \citenamefont {Zhou}, \citenamefont {Hao}, \citenamefont {Lu}, \citenamefont {Wang},\ and\ \citenamefont {Wen}}]{HHWen2023}%
  \BibitemOpen
  \bibfield  {author} {\bibinfo {author} {\bibfnamefont {Z.}~\bibnamefont {Liu}}, \bibinfo {author} {\bibfnamefont {M.}~\bibnamefont {Huo}}, \bibinfo {author} {\bibfnamefont {J.}~\bibnamefont {Li}}, \bibinfo {author} {\bibfnamefont {Q.}~\bibnamefont {Li}}, \bibinfo {author} {\bibfnamefont {Y.}~\bibnamefont {Liu}}, \bibinfo {author} {\bibfnamefont {Y.}~\bibnamefont {Dai}}, \bibinfo {author} {\bibfnamefont {X.}~\bibnamefont {Zhou}}, \bibinfo {author} {\bibfnamefont {J.}~\bibnamefont {Hao}}, \bibinfo {author} {\bibfnamefont {Y.}~\bibnamefont {Lu}}, \bibinfo {author} {\bibfnamefont {M.}~\bibnamefont {Wang}},\ and\ \bibinfo {author} {\bibfnamefont {H.-H.}\ \bibnamefont {Wen}},\ }\bibfield  {title} {\bibinfo {title} {{Electronic correlations and partial gap in the bilayer nickelate {La$_3$Ni$_2$O$_7$}}},\ }\href {https://doi.org/10.1038/s41467-024-52001-5} {\bibfield  {journal} {\bibinfo  {journal} {Nature Communications}\ }\textbf {\bibinfo {volume} {15}},\ \bibinfo {pages} {7570} (\bibinfo {year} {2024})}\BibitemShut {NoStop}%
\bibitem [{\citenamefont {Ko}\ \emph {et~al.}(2025)\citenamefont {Ko}, \citenamefont {Yu}, \citenamefont {Liu}, \citenamefont {Bhatt}, \citenamefont {Li}, \citenamefont {Thampy}, \citenamefont {Kuo}, \citenamefont {Wang}, \citenamefont {Lee}, \citenamefont {Lee}, \citenamefont {Lee}, \citenamefont {Goodge}, \citenamefont {Muller},\ and\ \citenamefont {Hwang}}]{ko2024signatures}%
  \BibitemOpen
  \bibfield  {author} {\bibinfo {author} {\bibfnamefont {E.~K.}\ \bibnamefont {Ko}}, \bibinfo {author} {\bibfnamefont {Y.}~\bibnamefont {Yu}}, \bibinfo {author} {\bibfnamefont {Y.}~\bibnamefont {Liu}}, \bibinfo {author} {\bibfnamefont {L.}~\bibnamefont {Bhatt}}, \bibinfo {author} {\bibfnamefont {J.}~\bibnamefont {Li}}, \bibinfo {author} {\bibfnamefont {V.}~\bibnamefont {Thampy}}, \bibinfo {author} {\bibfnamefont {C.-T.}\ \bibnamefont {Kuo}}, \bibinfo {author} {\bibfnamefont {B.~Y.}\ \bibnamefont {Wang}}, \bibinfo {author} {\bibfnamefont {Y.}~\bibnamefont {Lee}}, \bibinfo {author} {\bibfnamefont {K.}~\bibnamefont {Lee}}, \bibinfo {author} {\bibfnamefont {J.-S.}\ \bibnamefont {Lee}}, \bibinfo {author} {\bibfnamefont {B.~H.}\ \bibnamefont {Goodge}}, \bibinfo {author} {\bibfnamefont {D.~A.}\ \bibnamefont {Muller}},\ and\ \bibinfo {author} {\bibfnamefont {H.~Y.}\ \bibnamefont {Hwang}},\ }\bibfield  {title} {\bibinfo {title} {{Signatures of ambient pressure superconductivity in thin film {La$_3$Ni$_2$O$_7$}}},\ }\href {https://doi.org/10.1038/s41586-024-08525-3} {\bibfield  {journal} {\bibinfo  {journal} {Nature}\ }\textbf {\bibinfo {volume} {638}},\ \bibinfo {pages} {935} (\bibinfo {year} {2025})}\BibitemShut {NoStop}%
\bibitem [{\citenamefont {Zhou}\ \emph {et~al.}(2025)\citenamefont {Zhou}, \citenamefont {Lv}, \citenamefont {Wang}, \citenamefont {Nie}, \citenamefont {Chen}, \citenamefont {Li}, \citenamefont {Huang}, \citenamefont {Chen}, \citenamefont {Sun}, \citenamefont {Xue},\ and\ \citenamefont {Chen}}]{zhou2024ambientpressuresuperconductivityonset40}%
  \BibitemOpen
  \bibfield  {author} {\bibinfo {author} {\bibfnamefont {G.}~\bibnamefont {Zhou}}, \bibinfo {author} {\bibfnamefont {W.}~\bibnamefont {Lv}}, \bibinfo {author} {\bibfnamefont {H.}~\bibnamefont {Wang}}, \bibinfo {author} {\bibfnamefont {Z.}~\bibnamefont {Nie}}, \bibinfo {author} {\bibfnamefont {Y.}~\bibnamefont {Chen}}, \bibinfo {author} {\bibfnamefont {Y.}~\bibnamefont {Li}}, \bibinfo {author} {\bibfnamefont {H.}~\bibnamefont {Huang}}, \bibinfo {author} {\bibfnamefont {W.-Q.}\ \bibnamefont {Chen}}, \bibinfo {author} {\bibfnamefont {Y.-J.}\ \bibnamefont {Sun}}, \bibinfo {author} {\bibfnamefont {Q.-K.}\ \bibnamefont {Xue}},\ and\ \bibinfo {author} {\bibfnamefont {Z.}~\bibnamefont {Chen}},\ }\bibfield  {title} {\bibinfo {title} {{Ambient-pressure superconductivity onset above 40 K in (La,Pr)$_3$Ni$_2$O$_7$ films}},\ }\href {https://doi.org/10.1038/s41586-025-08755-z} {\bibfield  {journal} {\bibinfo  {journal} {Nature}\ }\textbf {\bibinfo {volume} {640}},\ \bibinfo {pages} {641} (\bibinfo {year} {2025})}\BibitemShut {NoStop}%
\bibitem [{\citenamefont {Liu}\ \emph {et~al.}(2025)\citenamefont {Liu}, \citenamefont {Ko}, \citenamefont {Tarn}, \citenamefont {Bhatt}, \citenamefont {Goodge}, \citenamefont {Muller}, \citenamefont {Raghu}, \citenamefont {Yu},\ and\ \citenamefont {Hwang}}]{liu2025superconductivitynormalstatetransportcompressively}%
  \BibitemOpen
  \bibfield  {author} {\bibinfo {author} {\bibfnamefont {Y.}~\bibnamefont {Liu}}, \bibinfo {author} {\bibfnamefont {E.~K.}\ \bibnamefont {Ko}}, \bibinfo {author} {\bibfnamefont {Y.}~\bibnamefont {Tarn}}, \bibinfo {author} {\bibfnamefont {L.}~\bibnamefont {Bhatt}}, \bibinfo {author} {\bibfnamefont {B.~H.}\ \bibnamefont {Goodge}}, \bibinfo {author} {\bibfnamefont {D.~A.}\ \bibnamefont {Muller}}, \bibinfo {author} {\bibfnamefont {S.}~\bibnamefont {Raghu}}, \bibinfo {author} {\bibfnamefont {Y.}~\bibnamefont {Yu}},\ and\ \bibinfo {author} {\bibfnamefont {H.~Y.}\ \bibnamefont {Hwang}},\ }\bibfield  {title} {\bibinfo {title} {{Superconductivity and normal-state transport in compressively strained La$_2$PrNi$_2$O$_7$ thin films}},\ }\href@noop {} {\bibfield  {journal} {\bibinfo  {journal} {arXiv:2501.08022}\ } (\bibinfo {year} {2025})}\BibitemShut {NoStop}%
\bibitem [{\citenamefont {Bhatt}\ \emph {et~al.}(2025)\citenamefont {Bhatt}, \citenamefont {Jiang}, \citenamefont {Ko}, \citenamefont {Schnitzer}, \citenamefont {Pan}, \citenamefont {Segedin}, \citenamefont {Liu}, \citenamefont {Yu}, \citenamefont {Zhao}, \citenamefont {Morales}, \citenamefont {Brooks}, \citenamefont {Botana}, \citenamefont {Hwang}, \citenamefont {Mundy}, \citenamefont {Muller},\ and\ \citenamefont {Goodge}}]{bhatt2025resolvingstructuraloriginssuperconductivity}%
  \BibitemOpen
  \bibfield  {author} {\bibinfo {author} {\bibfnamefont {L.}~\bibnamefont {Bhatt}}, \bibinfo {author} {\bibfnamefont {A.~Y.}\ \bibnamefont {Jiang}}, \bibinfo {author} {\bibfnamefont {E.~K.}\ \bibnamefont {Ko}}, \bibinfo {author} {\bibfnamefont {N.}~\bibnamefont {Schnitzer}}, \bibinfo {author} {\bibfnamefont {G.~A.}\ \bibnamefont {Pan}}, \bibinfo {author} {\bibfnamefont {D.~F.}\ \bibnamefont {Segedin}}, \bibinfo {author} {\bibfnamefont {Y.}~\bibnamefont {Liu}}, \bibinfo {author} {\bibfnamefont {Y.}~\bibnamefont {Yu}}, \bibinfo {author} {\bibfnamefont {Y.-F.}\ \bibnamefont {Zhao}}, \bibinfo {author} {\bibfnamefont {E.~A.}\ \bibnamefont {Morales}}, \bibinfo {author} {\bibfnamefont {C.~M.}\ \bibnamefont {Brooks}}, \bibinfo {author} {\bibfnamefont {A.~S.}\ \bibnamefont {Botana}}, \bibinfo {author} {\bibfnamefont {H.~Y.}\ \bibnamefont {Hwang}}, \bibinfo {author} {\bibfnamefont {J.~A.}\ \bibnamefont {Mundy}}, \bibinfo {author} {\bibfnamefont {D.~A.}\ \bibnamefont {Muller}},\ and\ \bibinfo {author} {\bibfnamefont {B.~H.}\ \bibnamefont {Goodge}},\ }\bibfield  {title} {\bibinfo {title} {{Resolving Structural Origins for Superconductivity in Strain-Engineered La$_3$Ni$_2$O$_7$ Thin Films}},\ }\href@noop {} {\bibfield  {journal} {\bibinfo  {journal} {arXiv:2501.08204}\ } (\bibinfo {year} {2025})}\BibitemShut {NoStop}%
\bibitem [{\citenamefont {{Li}}\ \emph {et~al.}(2025)\citenamefont {{Li}}, \citenamefont {{Zhou}}, \citenamefont {{Lv}}, \citenamefont {{Li}}, \citenamefont {{Yue}}, \citenamefont {{Huang}}, \citenamefont {{Xu}}, \citenamefont {{Shen}}, \citenamefont {{Miao}}, \citenamefont {{Song}}, \citenamefont {{Nie}}, \citenamefont {{Chen}}, \citenamefont {{Wang}}, \citenamefont {{Chen}}, \citenamefont {{Huang}}, \citenamefont {{Chen}}, \citenamefont {{Qian}}, \citenamefont {{Lin}}, \citenamefont {{He}}, \citenamefont {{Sun}}, \citenamefont {{Chen}},\ and\ \citenamefont {{Xue}}}]{ChenZY_ARPES2025}%
  \BibitemOpen
  \bibfield  {author} {\bibinfo {author} {\bibfnamefont {P.}~\bibnamefont {{Li}}}, \bibinfo {author} {\bibfnamefont {G.}~\bibnamefont {{Zhou}}}, \bibinfo {author} {\bibfnamefont {W.}~\bibnamefont {{Lv}}}, \bibinfo {author} {\bibfnamefont {Y.}~\bibnamefont {{Li}}}, \bibinfo {author} {\bibfnamefont {C.}~\bibnamefont {{Yue}}}, \bibinfo {author} {\bibfnamefont {H.}~\bibnamefont {{Huang}}}, \bibinfo {author} {\bibfnamefont {L.}~\bibnamefont {{Xu}}}, \bibinfo {author} {\bibfnamefont {J.}~\bibnamefont {{Shen}}}, \bibinfo {author} {\bibfnamefont {Y.}~\bibnamefont {{Miao}}}, \bibinfo {author} {\bibfnamefont {W.}~\bibnamefont {{Song}}}, \bibinfo {author} {\bibfnamefont {Z.}~\bibnamefont {{Nie}}}, \bibinfo {author} {\bibfnamefont {Y.}~\bibnamefont {{Chen}}}, \bibinfo {author} {\bibfnamefont {H.}~\bibnamefont {{Wang}}}, \bibinfo {author} {\bibfnamefont {W.}~\bibnamefont {{Chen}}}, \bibinfo {author} {\bibfnamefont {Y.}~\bibnamefont {{Huang}}}, \bibinfo {author} {\bibfnamefont {Z.-H.}\ \bibnamefont {{Chen}}}, \bibinfo {author} {\bibfnamefont {T.}~\bibnamefont {{Qian}}}, \bibinfo {author} {\bibfnamefont {J.}~\bibnamefont {{Lin}}}, \bibinfo {author} {\bibfnamefont {J.}~\bibnamefont {{He}}}, \bibinfo {author} {\bibfnamefont {Y.-J.}\ \bibnamefont {{Sun}}}, \bibinfo {author} {\bibfnamefont {Z.}~\bibnamefont {{Chen}}},\ and\ \bibinfo {author} {\bibfnamefont {Q.-K.}\ \bibnamefont {{Xue}}},\ }\bibfield  {title} {\bibinfo {title} {{Angle-resolved photoemission spectroscopy of superconducting (La,Pr)$_3$Ni$_2$O$_7$/SrLaAlO$_4$ heterostructures}},\ }\bibfield  {journal} {\bibinfo  {journal} {arXiv e-prints}\ }\href {https://doi.org/10.48550/arXiv.2501.09255} {10.48550/arXiv.2501.09255} (\bibinfo {year} {2025})\BibitemShut {NoStop}%
\bibitem [{\citenamefont {{Wang}}\ \emph {et~al.}(2025{\natexlab{a}})\citenamefont {{Wang}}, \citenamefont {{Zhong}}, \citenamefont {{Abadi}}, \citenamefont {{Liu}}, \citenamefont {{Yu}}, \citenamefont {{Zhang}}, \citenamefont {{Wu}}, \citenamefont {{Wang}}, \citenamefont {{Li}}, \citenamefont {{Tarn}}, \citenamefont {{Kyo Ko}}, \citenamefont {{Thampy}}, \citenamefont {{Hashimoto}}, \citenamefont {{Lu}}, \citenamefont {{Lee}}, \citenamefont {{Devereaux}}, \citenamefont {{Jia}}, \citenamefont {{Hwang}},\ and\ \citenamefont {{Shen}}}]{ShenZX2025}%
  \BibitemOpen
  \bibfield  {author} {\bibinfo {author} {\bibfnamefont {B.~Y.}\ \bibnamefont {{Wang}}}, \bibinfo {author} {\bibfnamefont {Y.}~\bibnamefont {{Zhong}}}, \bibinfo {author} {\bibfnamefont {S.}~\bibnamefont {{Abadi}}}, \bibinfo {author} {\bibfnamefont {Y.}~\bibnamefont {{Liu}}}, \bibinfo {author} {\bibfnamefont {Y.}~\bibnamefont {{Yu}}}, \bibinfo {author} {\bibfnamefont {X.}~\bibnamefont {{Zhang}}}, \bibinfo {author} {\bibfnamefont {Y.-M.}\ \bibnamefont {{Wu}}}, \bibinfo {author} {\bibfnamefont {R.}~\bibnamefont {{Wang}}}, \bibinfo {author} {\bibfnamefont {J.}~\bibnamefont {{Li}}}, \bibinfo {author} {\bibfnamefont {Y.}~\bibnamefont {{Tarn}}}, \bibinfo {author} {\bibfnamefont {E.}~\bibnamefont {{Kyo Ko}}}, \bibinfo {author} {\bibfnamefont {V.}~\bibnamefont {{Thampy}}}, \bibinfo {author} {\bibfnamefont {M.}~\bibnamefont {{Hashimoto}}}, \bibinfo {author} {\bibfnamefont {D.}~\bibnamefont {{Lu}}}, \bibinfo {author} {\bibfnamefont {Y.~S.}\ \bibnamefont {{Lee}}}, \bibinfo {author} {\bibfnamefont {T.~P.}\ \bibnamefont {{Devereaux}}}, \bibinfo {author} {\bibfnamefont {C.}~\bibnamefont {{Jia}}}, \bibinfo {author} {\bibfnamefont {H.~Y.}\ \bibnamefont {{Hwang}}},\ and\ \bibinfo {author} {\bibfnamefont {Z.-X.}\ \bibnamefont {{Shen}}},\ }\bibfield  {title} {\bibinfo {title} {{Electronic structure of compressively strained thin film La$_2$PrNi$_2$O$_7$}},\ }\bibfield  {journal} {\bibinfo  {journal} {arXiv e-prints}\ }\href {https://doi.org/10.48550/arXiv.2504.16372} {10.48550/arXiv.2504.16372} (\bibinfo {year} {2025}{\natexlab{a}})\BibitemShut {NoStop}%
\bibitem [{\citenamefont {Yang}\ \emph {et~al.}(2023{\natexlab{a}})\citenamefont {Yang}, \citenamefont {Wang},\ and\ \citenamefont {Wang}}]{Wang327prb}%
  \BibitemOpen
  \bibfield  {author} {\bibinfo {author} {\bibfnamefont {Q.-G.}\ \bibnamefont {Yang}}, \bibinfo {author} {\bibfnamefont {D.}~\bibnamefont {Wang}},\ and\ \bibinfo {author} {\bibfnamefont {Q.-H.}\ \bibnamefont {Wang}},\ }\bibfield  {title} {\bibinfo {title} {{Possible ${s}_{\ifmmode\pm\else\textpm\fi{}}$-wave superconductivity in ${\mathrm{La}}_{3}{\mathrm{Ni}}_{2}{\mathrm{O}}_{7}$}},\ }\href {https://doi.org/10.1103/PhysRevB.108.L140505} {\bibfield  {journal} {\bibinfo  {journal} {Phys. Rev. B}\ }\textbf {\bibinfo {volume} {108}},\ \bibinfo {pages} {L140505} (\bibinfo {year} {2023}{\natexlab{a}})}\BibitemShut {NoStop}%
\bibitem [{\citenamefont {Lu}\ \emph {et~al.}(2024{\natexlab{a}})\citenamefont {Lu}, \citenamefont {Pan}, \citenamefont {Yang},\ and\ \citenamefont {Wu}}]{lu2024interlayer}%
  \BibitemOpen
  \bibfield  {author} {\bibinfo {author} {\bibfnamefont {C.}~\bibnamefont {Lu}}, \bibinfo {author} {\bibfnamefont {Z.}~\bibnamefont {Pan}}, \bibinfo {author} {\bibfnamefont {F.}~\bibnamefont {Yang}},\ and\ \bibinfo {author} {\bibfnamefont {C.}~\bibnamefont {Wu}},\ }\bibfield  {title} {\bibinfo {title} {{Interlayer-Coupling-Driven High-Temperature Superconductivity in ${\mathrm{La}}_{3}{\mathrm{Ni}}_{2}{\mathrm{O}}_{7}$ under Pressure}},\ }\href {https://doi.org/10.1103/PhysRevLett.132.146002} {\bibfield  {journal} {\bibinfo  {journal} {Phys. Rev. Lett.}\ }\textbf {\bibinfo {volume} {132}},\ \bibinfo {pages} {146002} (\bibinfo {year} {2024}{\natexlab{a}})}\BibitemShut {NoStop}%
\bibitem [{\citenamefont {Oh}\ and\ \citenamefont {Zhang}(2023{\natexlab{a}})}]{HYZhangtype2}%
  \BibitemOpen
  \bibfield  {author} {\bibinfo {author} {\bibfnamefont {H.}~\bibnamefont {Oh}}\ and\ \bibinfo {author} {\bibfnamefont {Y.-H.}\ \bibnamefont {Zhang}},\ }\bibfield  {title} {\bibinfo {title} {{Type-II $t\ensuremath{-}J$ model and shared superexchange coupling from Hund's rule in superconducting ${\mathrm{La}}_{3}{\mathrm{Ni}}_{2}{\mathrm{O}}_{7}$}},\ }\href {https://doi.org/10.1103/PhysRevB.108.174511} {\bibfield  {journal} {\bibinfo  {journal} {Phys. Rev. B}\ }\textbf {\bibinfo {volume} {108}},\ \bibinfo {pages} {174511} (\bibinfo {year} {2023}{\natexlab{a}})}\BibitemShut {NoStop}%
\bibitem [{\citenamefont {Liu}\ \emph {et~al.}(2023{\natexlab{a}})\citenamefont {Liu}, \citenamefont {Mei}, \citenamefont {Ye}, \citenamefont {Chen},\ and\ \citenamefont {Yang}}]{FangYang327prl}%
  \BibitemOpen
  \bibfield  {author} {\bibinfo {author} {\bibfnamefont {Y.-B.}\ \bibnamefont {Liu}}, \bibinfo {author} {\bibfnamefont {J.-W.}\ \bibnamefont {Mei}}, \bibinfo {author} {\bibfnamefont {F.}~\bibnamefont {Ye}}, \bibinfo {author} {\bibfnamefont {W.-Q.}\ \bibnamefont {Chen}},\ and\ \bibinfo {author} {\bibfnamefont {F.}~\bibnamefont {Yang}},\ }\bibfield  {title} {\bibinfo {title} {{${s}^{\ifmmode\pm\else\textpm\fi{}}$-Wave Pairing and the Destructive Role of Apical-Oxygen Deficiencies in ${\mathrm{La}}_{3}{\mathrm{Ni}}_{2}{\mathrm{O}}_{7}$ under Pressure}},\ }\href {https://doi.org/10.1103/PhysRevLett.131.236002} {\bibfield  {journal} {\bibinfo  {journal} {Phys. Rev. Lett.}\ }\textbf {\bibinfo {volume} {131}},\ \bibinfo {pages} {236002} (\bibinfo {year} {2023}{\natexlab{a}})}\BibitemShut {NoStop}%
\bibitem [{\citenamefont {Qu}\ \emph {et~al.}(2024{\natexlab{a}})\citenamefont {Qu}, \citenamefont {Qu}, \citenamefont {Chen}, \citenamefont {Wu}, \citenamefont {Yang}, \citenamefont {Li},\ and\ \citenamefont {Su}}]{WeiLi327prl}%
  \BibitemOpen
  \bibfield  {author} {\bibinfo {author} {\bibfnamefont {X.-Z.}\ \bibnamefont {Qu}}, \bibinfo {author} {\bibfnamefont {D.-W.}\ \bibnamefont {Qu}}, \bibinfo {author} {\bibfnamefont {J.}~\bibnamefont {Chen}}, \bibinfo {author} {\bibfnamefont {C.}~\bibnamefont {Wu}}, \bibinfo {author} {\bibfnamefont {F.}~\bibnamefont {Yang}}, \bibinfo {author} {\bibfnamefont {W.}~\bibnamefont {Li}},\ and\ \bibinfo {author} {\bibfnamefont {G.}~\bibnamefont {Su}},\ }\bibfield  {title} {\bibinfo {title} {{Bilayer ${t\text{\ensuremath{-}}J\text{\ensuremath{-}}J}_{\ensuremath{\perp}}$ Model and Magnetically Mediated Pairing in the Pressurized Nickelate ${\mathrm{La}}_{3}{\mathrm{Ni}}_{2}{\mathrm{O}}_{7}$}},\ }\href {https://doi.org/10.1103/PhysRevLett.132.036502} {\bibfield  {journal} {\bibinfo  {journal} {Phys. Rev. Lett.}\ }\textbf {\bibinfo {volume} {132}},\ \bibinfo {pages} {036502} (\bibinfo {year} {2024}{\natexlab{a}})}\BibitemShut {NoStop}%
\bibitem [{\citenamefont {Yang}\ \emph {et~al.}(2023{\natexlab{b}})\citenamefont {Yang}, \citenamefont {Zhang},\ and\ \citenamefont {Zhang}}]{YifengYang327prb}%
  \BibitemOpen
  \bibfield  {author} {\bibinfo {author} {\bibfnamefont {Y.-f.}\ \bibnamefont {Yang}}, \bibinfo {author} {\bibfnamefont {G.-M.}\ \bibnamefont {Zhang}},\ and\ \bibinfo {author} {\bibfnamefont {F.-C.}\ \bibnamefont {Zhang}},\ }\bibfield  {title} {\bibinfo {title} {{Interlayer valence bonds and two-component theory for high-${T}_{c}$ superconductivity of ${\mathrm{La}}_{3}{\mathrm{Ni}}_{2}{\mathrm{O}}_{7}$ under pressure}},\ }\href {https://doi.org/10.1103/PhysRevB.108.L201108} {\bibfield  {journal} {\bibinfo  {journal} {Phys. Rev. B}\ }\textbf {\bibinfo {volume} {108}},\ \bibinfo {pages} {L201108} (\bibinfo {year} {2023}{\natexlab{b}})}\BibitemShut {NoStop}%
\bibitem [{\citenamefont {Qin}\ and\ \citenamefont {Yang}(2023)}]{YifengYang327prb2}%
  \BibitemOpen
  \bibfield  {author} {\bibinfo {author} {\bibfnamefont {Q.}~\bibnamefont {Qin}}\ and\ \bibinfo {author} {\bibfnamefont {Y.-f.}\ \bibnamefont {Yang}},\ }\bibfield  {title} {\bibinfo {title} {{High-${T}_{c}$ superconductivity by mobilizing local spin singlets and possible route to higher ${T}_{c}$ in pressurized ${\mathrm{La}}_{3}{\mathrm{Ni}}_{2}{\mathrm{O}}_{7}$}},\ }\href {https://doi.org/10.1103/PhysRevB.108.L140504} {\bibfield  {journal} {\bibinfo  {journal} {Phys. Rev. B}\ }\textbf {\bibinfo {volume} {108}},\ \bibinfo {pages} {L140504} (\bibinfo {year} {2023})}\BibitemShut {NoStop}%
\bibitem [{\citenamefont {Lu}\ \emph {et~al.}(2023)\citenamefont {Lu}, \citenamefont {Li}, \citenamefont {Zeng}, \citenamefont {Hou}, \citenamefont {Wang}, \citenamefont {Yang},\ and\ \citenamefont {You}}]{YiZhuangYouSMG}%
  \BibitemOpen
  \bibfield  {author} {\bibinfo {author} {\bibfnamefont {D.-C.}\ \bibnamefont {Lu}}, \bibinfo {author} {\bibfnamefont {M.}~\bibnamefont {Li}}, \bibinfo {author} {\bibfnamefont {Z.-Y.}\ \bibnamefont {Zeng}}, \bibinfo {author} {\bibfnamefont {W.}~\bibnamefont {Hou}}, \bibinfo {author} {\bibfnamefont {J.}~\bibnamefont {Wang}}, \bibinfo {author} {\bibfnamefont {F.}~\bibnamefont {Yang}},\ and\ \bibinfo {author} {\bibfnamefont {Y.-Z.}\ \bibnamefont {You}},\ }\bibfield  {title} {\bibinfo {title} {{Superconductivity from Doping Symmetric Mass Generation Insulators: Application to La$_3$Ni$_2$O$_7$ under Pressure}},\ }\href@noop {} {\bibfield  {journal} {\bibinfo  {journal} {arXiv: 2308.11195}\ } (\bibinfo {year} {2023})}\BibitemShut {NoStop}%
\bibitem [{\citenamefont {Tian}\ \emph {et~al.}(2024{\natexlab{a}})\citenamefont {Tian}, \citenamefont {Chen}, \citenamefont {Wang}, \citenamefont {He},\ and\ \citenamefont {Lu}}]{tian2023correlation}%
  \BibitemOpen
  \bibfield  {author} {\bibinfo {author} {\bibfnamefont {Y.-H.}\ \bibnamefont {Tian}}, \bibinfo {author} {\bibfnamefont {Y.}~\bibnamefont {Chen}}, \bibinfo {author} {\bibfnamefont {J.-M.}\ \bibnamefont {Wang}}, \bibinfo {author} {\bibfnamefont {R.-Q.}\ \bibnamefont {He}},\ and\ \bibinfo {author} {\bibfnamefont {Z.-Y.}\ \bibnamefont {Lu}},\ }\bibfield  {title} {\bibinfo {title} {{Correlation effects and concomitant two-orbital ${s}_{\ifmmode\pm\else\textpm\fi{}}$-wave superconductivity in ${\mathrm{La}}_{3}{\mathrm{Ni}}_{2}{\mathrm{O}}_{7}$ under high pressure}},\ }\href {https://doi.org/10.1103/PhysRevB.109.165154} {\bibfield  {journal} {\bibinfo  {journal} {Phys. Rev. B}\ }\textbf {\bibinfo {volume} {109}},\ \bibinfo {pages} {165154} (\bibinfo {year} {2024}{\natexlab{a}})}\BibitemShut {NoStop}%
\bibitem [{\citenamefont {Zhang}\ \emph {et~al.}(2023{\natexlab{b}})\citenamefont {Zhang}, \citenamefont {Lin}, \citenamefont {Moreo}, \citenamefont {Maier},\ and\ \citenamefont {Dagotto}}]{Dagotto327prb}%
  \BibitemOpen
  \bibfield  {author} {\bibinfo {author} {\bibfnamefont {Y.}~\bibnamefont {Zhang}}, \bibinfo {author} {\bibfnamefont {L.-F.}\ \bibnamefont {Lin}}, \bibinfo {author} {\bibfnamefont {A.}~\bibnamefont {Moreo}}, \bibinfo {author} {\bibfnamefont {T.~A.}\ \bibnamefont {Maier}},\ and\ \bibinfo {author} {\bibfnamefont {E.}~\bibnamefont {Dagotto}},\ }\bibfield  {title} {\bibinfo {title} {{Trends in electronic structures and ${s}_{\ifmmode\pm\else\textpm\fi{}}$-wave pairing for the rare-earth series in bilayer nickelate superconductor ${R}_{3}{\mathrm{Ni}}_{2}{\mathrm{O}}_{7}$}},\ }\href {https://doi.org/10.1103/PhysRevB.108.165141} {\bibfield  {journal} {\bibinfo  {journal} {Phys. Rev. B}\ }\textbf {\bibinfo {volume} {108}},\ \bibinfo {pages} {165141} (\bibinfo {year} {2023}{\natexlab{b}})}\BibitemShut {NoStop}%
\bibitem [{\citenamefont {Zhang}\ \emph {et~al.}(2024{\natexlab{a}})\citenamefont {Zhang}, \citenamefont {Lin}, \citenamefont {Moreo}, \citenamefont {Maier},\ and\ \citenamefont {Dagotto}}]{zhang2024structural}%
  \BibitemOpen
  \bibfield  {author} {\bibinfo {author} {\bibfnamefont {Y.}~\bibnamefont {Zhang}}, \bibinfo {author} {\bibfnamefont {L.-F.}\ \bibnamefont {Lin}}, \bibinfo {author} {\bibfnamefont {A.}~\bibnamefont {Moreo}}, \bibinfo {author} {\bibfnamefont {T.~A.}\ \bibnamefont {Maier}},\ and\ \bibinfo {author} {\bibfnamefont {E.}~\bibnamefont {Dagotto}},\ }\bibfield  {title} {\bibinfo {title} {{Structural phase transition, s±-wave pairing, and magnetic stripe order in bilayered superconductor {La$_3$Ni$_2$O$_7$} under pressure}},\ }\href {https://doi.org/10.1038/s41467-024-46622-z} {\bibfield  {journal} {\bibinfo  {journal} {Nature Communications}\ }\textbf {\bibinfo {volume} {15}},\ \bibinfo {pages} {2470} (\bibinfo {year} {2024}{\natexlab{a}})}\BibitemShut {NoStop}%
\bibitem [{\citenamefont {Jiang}\ \emph {et~al.}(2024{\natexlab{a}})\citenamefont {Jiang}, \citenamefont {Wang},\ and\ \citenamefont {Zhang}}]{Jiang_2024}%
  \BibitemOpen
  \bibfield  {author} {\bibinfo {author} {\bibfnamefont {K.}~\bibnamefont {Jiang}}, \bibinfo {author} {\bibfnamefont {Z.}~\bibnamefont {Wang}},\ and\ \bibinfo {author} {\bibfnamefont {F.-C.}\ \bibnamefont {Zhang}},\ }\bibfield  {title} {\bibinfo {title} {{High-Temperature Superconductivity in La$_3$Ni$_2$O$_7$}},\ }\href {https://doi.org/10.1088/0256-307X/41/1/017402} {\bibfield  {journal} {\bibinfo  {journal} {Chinese Physics Letters}\ }\textbf {\bibinfo {volume} {41}},\ \bibinfo {pages} {017402} (\bibinfo {year} {2024}{\natexlab{a}})}\BibitemShut {NoStop}%
\bibitem [{\citenamefont {Lechermann}\ \emph {et~al.}(2023{\natexlab{b}})\citenamefont {Lechermann}, \citenamefont {Gondolf}, \citenamefont {B\"otzel},\ and\ \citenamefont {Eremin}}]{PhysRevB.108.L201121}%
  \BibitemOpen
  \bibfield  {author} {\bibinfo {author} {\bibfnamefont {F.}~\bibnamefont {Lechermann}}, \bibinfo {author} {\bibfnamefont {J.}~\bibnamefont {Gondolf}}, \bibinfo {author} {\bibfnamefont {S.}~\bibnamefont {B\"otzel}},\ and\ \bibinfo {author} {\bibfnamefont {I.~M.}\ \bibnamefont {Eremin}},\ }\bibfield  {title} {\bibinfo {title} {{Electronic correlations and superconducting instability in ${\mathrm{La}}_{3}{\mathrm{Ni}}_{2}{\mathrm{O}}_{7}$ under high pressure}},\ }\href {https://doi.org/10.1103/PhysRevB.108.L201121} {\bibfield  {journal} {\bibinfo  {journal} {Phys. Rev. B}\ }\textbf {\bibinfo {volume} {108}},\ \bibinfo {pages} {L201121} (\bibinfo {year} {2023}{\natexlab{b}})}\BibitemShut {NoStop}%
\bibitem [{\citenamefont {Liao}\ \emph {et~al.}(2023{\natexlab{a}})\citenamefont {Liao}, \citenamefont {Chen}, \citenamefont {Duan}, \citenamefont {Wang}, \citenamefont {Liu}, \citenamefont {Yu},\ and\ \citenamefont {Si}}]{liao2023electron}%
  \BibitemOpen
  \bibfield  {author} {\bibinfo {author} {\bibfnamefont {Z.}~\bibnamefont {Liao}}, \bibinfo {author} {\bibfnamefont {L.}~\bibnamefont {Chen}}, \bibinfo {author} {\bibfnamefont {G.}~\bibnamefont {Duan}}, \bibinfo {author} {\bibfnamefont {Y.}~\bibnamefont {Wang}}, \bibinfo {author} {\bibfnamefont {C.}~\bibnamefont {Liu}}, \bibinfo {author} {\bibfnamefont {R.}~\bibnamefont {Yu}},\ and\ \bibinfo {author} {\bibfnamefont {Q.}~\bibnamefont {Si}},\ }\bibfield  {title} {\bibinfo {title} {{Electron correlations and superconductivity in ${\mathrm{La}}_{3}{\mathrm{Ni}}_{2}{\mathrm{O}}_{7}$ under pressure tuning}},\ }\href {https://doi.org/10.1103/PhysRevB.108.214522} {\bibfield  {journal} {\bibinfo  {journal} {Phys. Rev. B}\ }\textbf {\bibinfo {volume} {108}},\ \bibinfo {pages} {214522} (\bibinfo {year} {2023}{\natexlab{a}})}\BibitemShut {NoStop}%
\bibitem [{\citenamefont {Ryee}\ \emph {et~al.}(2024)\citenamefont {Ryee}, \citenamefont {Witt},\ and\ \citenamefont {Wehling}}]{ryee2024quenched}%
  \BibitemOpen
  \bibfield  {author} {\bibinfo {author} {\bibfnamefont {S.}~\bibnamefont {Ryee}}, \bibinfo {author} {\bibfnamefont {N.}~\bibnamefont {Witt}},\ and\ \bibinfo {author} {\bibfnamefont {T.~O.}\ \bibnamefont {Wehling}},\ }\bibfield  {title} {\bibinfo {title} {{Quenched Pair Breaking by Interlayer Correlations as a Key to Superconductivity in ${\mathrm{La}}_{3}{\mathrm{Ni}}_{2}{\mathrm{O}}_{7}$}},\ }\href {https://doi.org/10.1103/PhysRevLett.133.096002} {\bibfield  {journal} {\bibinfo  {journal} {Phys. Rev. Lett.}\ }\textbf {\bibinfo {volume} {133}},\ \bibinfo {pages} {096002} (\bibinfo {year} {2024})}\BibitemShut {NoStop}%
\bibitem [{\citenamefont {Luo}\ \emph {et~al.}(2024)\citenamefont {Luo}, \citenamefont {Lv}, \citenamefont {Wang}, \citenamefont {Wú},\ and\ \citenamefont {Yao}}]{luo2023hightc}%
  \BibitemOpen
  \bibfield  {author} {\bibinfo {author} {\bibfnamefont {Z.}~\bibnamefont {Luo}}, \bibinfo {author} {\bibfnamefont {B.}~\bibnamefont {Lv}}, \bibinfo {author} {\bibfnamefont {M.}~\bibnamefont {Wang}}, \bibinfo {author} {\bibfnamefont {W.}~\bibnamefont {Wú}},\ and\ \bibinfo {author} {\bibfnamefont {D.-X.}\ \bibnamefont {Yao}},\ }\bibfield  {title} {\bibinfo {title} {{High-{$T_c$} superconductivity in {La$_3$Ni$_2$O$_7$} based on the bilayer two-orbital t-{J} model}},\ }\href {https://doi.org/10.1038/s41535-024-00668-w} {\bibfield  {journal} {\bibinfo  {journal} {npj Quantum Materials}\ }\textbf {\bibinfo {volume} {9}},\ \bibinfo {pages} {61} (\bibinfo {year} {2024})}\BibitemShut {NoStop}%
\bibitem [{\citenamefont {Jiang}\ \emph {et~al.}(2024{\natexlab{b}})\citenamefont {Jiang}, \citenamefont {Wang},\ and\ \citenamefont {Zhang}}]{KJiang:17402}%
  \BibitemOpen
  \bibfield  {author} {\bibinfo {author} {\bibfnamefont {K.}~\bibnamefont {Jiang}}, \bibinfo {author} {\bibfnamefont {Z.}~\bibnamefont {Wang}},\ and\ \bibinfo {author} {\bibfnamefont {F.-C.}\ \bibnamefont {Zhang}},\ }\bibfield  {title} {\bibinfo {title} {{High-Temperature Superconductivity in La$_3$Ni$_2$O$_7$}},\ }\href {https://doi.org/10.1088/0256-307X/41/1/017402} {\bibfield  {journal} {\bibinfo  {journal} {Chinese Physics Letters}\ }\textbf {\bibinfo {volume} {41}},\ \bibinfo {eid} {017402} (\bibinfo {year} {2024}{\natexlab{b}})}\BibitemShut {NoStop}%
\bibitem [{\citenamefont {Jiang}\ \emph {et~al.}(2024{\natexlab{c}})\citenamefont {Jiang}, \citenamefont {Hou}, \citenamefont {Fan}, \citenamefont {Lang},\ and\ \citenamefont {Ku}}]{KuWeiprl}%
  \BibitemOpen
  \bibfield  {author} {\bibinfo {author} {\bibfnamefont {R.}~\bibnamefont {Jiang}}, \bibinfo {author} {\bibfnamefont {J.}~\bibnamefont {Hou}}, \bibinfo {author} {\bibfnamefont {Z.}~\bibnamefont {Fan}}, \bibinfo {author} {\bibfnamefont {Z.-J.}\ \bibnamefont {Lang}},\ and\ \bibinfo {author} {\bibfnamefont {W.}~\bibnamefont {Ku}},\ }\bibfield  {title} {\bibinfo {title} {{Pressure Driven Fractionalization of Ionic Spins Results in Cupratelike High-${T}_{c}$ Superconductivity in ${\mathrm{La}}_{3}{\mathrm{Ni}}_{2}{\mathrm{O}}_{7}$}},\ }\href {https://doi.org/10.1103/PhysRevLett.132.126503} {\bibfield  {journal} {\bibinfo  {journal} {Phys. Rev. Lett.}\ }\textbf {\bibinfo {volume} {132}},\ \bibinfo {pages} {126503} (\bibinfo {year} {2024}{\natexlab{c}})}\BibitemShut {NoStop}%
\bibitem [{\citenamefont {Fan}\ \emph {et~al.}(2024{\natexlab{a}})\citenamefont {Fan}, \citenamefont {Zhang}, \citenamefont {Zhan}, \citenamefont {Lv}, \citenamefont {Jiang}, \citenamefont {Normand},\ and\ \citenamefont {Xiang}}]{fan2023superconductivity}%
  \BibitemOpen
  \bibfield  {author} {\bibinfo {author} {\bibfnamefont {Z.}~\bibnamefont {Fan}}, \bibinfo {author} {\bibfnamefont {J.-F.}\ \bibnamefont {Zhang}}, \bibinfo {author} {\bibfnamefont {B.}~\bibnamefont {Zhan}}, \bibinfo {author} {\bibfnamefont {D.}~\bibnamefont {Lv}}, \bibinfo {author} {\bibfnamefont {X.-Y.}\ \bibnamefont {Jiang}}, \bibinfo {author} {\bibfnamefont {B.}~\bibnamefont {Normand}},\ and\ \bibinfo {author} {\bibfnamefont {T.}~\bibnamefont {Xiang}},\ }\bibfield  {title} {\bibinfo {title} {{Superconductivity in nickelate and cuprate superconductors with strong bilayer coupling}},\ }\href {https://doi.org/10.1103/PhysRevB.110.024514} {\bibfield  {journal} {\bibinfo  {journal} {Phys. Rev. B}\ }\textbf {\bibinfo {volume} {110}},\ \bibinfo {pages} {024514} (\bibinfo {year} {2024}{\natexlab{a}})}\BibitemShut {NoStop}%
\bibitem [{\citenamefont {Zhan}\ \emph {et~al.}(2025)\citenamefont {Zhan}, \citenamefont {Gu}, \citenamefont {Wu},\ and\ \citenamefont {Hu}}]{zhan2024cooperation}%
  \BibitemOpen
  \bibfield  {author} {\bibinfo {author} {\bibfnamefont {J.}~\bibnamefont {Zhan}}, \bibinfo {author} {\bibfnamefont {Y.}~\bibnamefont {Gu}}, \bibinfo {author} {\bibfnamefont {X.}~\bibnamefont {Wu}},\ and\ \bibinfo {author} {\bibfnamefont {J.}~\bibnamefont {Hu}},\ }\bibfield  {title} {\bibinfo {title} {{Cooperation between Electron-Phonon Coupling and Electronic Interaction in Bilayer Nickelates ${\mathrm{La}}_{3}{\mathrm{Ni}}_{2}{\mathrm{O}}_{7}$}},\ }\href {https://doi.org/10.1103/PhysRevLett.134.136002} {\bibfield  {journal} {\bibinfo  {journal} {Phys. Rev. Lett.}\ }\textbf {\bibinfo {volume} {134}},\ \bibinfo {pages} {136002} (\bibinfo {year} {2025})}\BibitemShut {NoStop}%
\bibitem [{\citenamefont {Xia}\ \emph {et~al.}(2025{\natexlab{a}})\citenamefont {Xia}, \citenamefont {Liu}, \citenamefont {Zhou},\ and\ \citenamefont {Chen}}]{ChenHH2025}%
  \BibitemOpen
  \bibfield  {author} {\bibinfo {author} {\bibfnamefont {C.}~\bibnamefont {Xia}}, \bibinfo {author} {\bibfnamefont {H.}~\bibnamefont {Liu}}, \bibinfo {author} {\bibfnamefont {S.}~\bibnamefont {Zhou}},\ and\ \bibinfo {author} {\bibfnamefont {H.}~\bibnamefont {Chen}},\ }\bibfield  {title} {\bibinfo {title} {{Sensitive dependence of pairing symmetry on {Ni}-eg crystal field splitting in the nickelate superconductor {La$_3$Ni$_2$O$_7$}}},\ }\href {https://doi.org/10.1038/s41467-025-56206-0} {\bibfield  {journal} {\bibinfo  {journal} {Nature Communications}\ }\textbf {\bibinfo {volume} {16}},\ \bibinfo {pages} {1054} (\bibinfo {year} {2025}{\natexlab{a}})}\BibitemShut {NoStop}%
\bibitem [{\citenamefont {Bejas}\ \emph {et~al.}(2025)\citenamefont {Bejas}, \citenamefont {Wu}, \citenamefont {Chakraborty}, \citenamefont {Schnyder},\ and\ \citenamefont {Greco}}]{PhysRevB.111.144514}%
  \BibitemOpen
  \bibfield  {author} {\bibinfo {author} {\bibfnamefont {M.}~\bibnamefont {Bejas}}, \bibinfo {author} {\bibfnamefont {X.}~\bibnamefont {Wu}}, \bibinfo {author} {\bibfnamefont {D.}~\bibnamefont {Chakraborty}}, \bibinfo {author} {\bibfnamefont {A.~P.}\ \bibnamefont {Schnyder}},\ and\ \bibinfo {author} {\bibfnamefont {A.}~\bibnamefont {Greco}},\ }\bibfield  {title} {\bibinfo {title} {{Out-of-plane bond-order phase, superconductivity, and their competition in the $t$- ${J}_{\ensuremath{\parallel}}\text{\ensuremath{-}}{J}_{\ensuremath{\perp}}$ model: Possible implications for bilayer nickelates}},\ }\href {https://doi.org/10.1103/PhysRevB.111.144514} {\bibfield  {journal} {\bibinfo  {journal} {Phys. Rev. B}\ }\textbf {\bibinfo {volume} {111}},\ \bibinfo {pages} {144514} (\bibinfo {year} {2025})}\BibitemShut {NoStop}%
\bibitem [{\citenamefont {Xi}\ \emph {et~al.}(2025)\citenamefont {Xi}, \citenamefont {Yu},\ and\ \citenamefont {Li}}]{LiJX_nonlocal}%
  \BibitemOpen
  \bibfield  {author} {\bibinfo {author} {\bibfnamefont {W.}~\bibnamefont {Xi}}, \bibinfo {author} {\bibfnamefont {S.-L.}\ \bibnamefont {Yu}},\ and\ \bibinfo {author} {\bibfnamefont {J.-X.}\ \bibnamefont {Li}},\ }\bibfield  {title} {\bibinfo {title} {{Transition from ${s}_{\ifmmode\pm\else\textpm\fi{}}$-wave to ${d}_{{x}^{2}\ensuremath{-}{y}^{2}}$-wave superconductivity driven by interlayer interaction in the bilayer two-orbital model of ${\text{La}}_{3}{\text{Ni}}_{2}{\text{O}}_{7}$}},\ }\href {https://doi.org/10.1103/PhysRevB.111.104505} {\bibfield  {journal} {\bibinfo  {journal} {Phys. Rev. B}\ }\textbf {\bibinfo {volume} {111}},\ \bibinfo {pages} {104505} (\bibinfo {year} {2025})}\BibitemShut {NoStop}%
\bibitem [{\citenamefont {{Zhan}}\ \emph {et~al.}(2025)\citenamefont {{Zhan}}, \citenamefont {{Le}}, \citenamefont {{Wu}},\ and\ \citenamefont {{Hu}}}]{2025arXiv250318877Z}%
  \BibitemOpen
  \bibfield  {author} {\bibinfo {author} {\bibfnamefont {J.}~\bibnamefont {{Zhan}}}, \bibinfo {author} {\bibfnamefont {C.}~\bibnamefont {{Le}}}, \bibinfo {author} {\bibfnamefont {X.}~\bibnamefont {{Wu}}},\ and\ \bibinfo {author} {\bibfnamefont {J.}~\bibnamefont {{Hu}}},\ }\bibfield  {title} {\bibinfo {title} {{Impact of Nonlocal Coulomb Repulsion on Superconductivity and Density-Wave Orders in Bilayer Nickelates}},\ }\bibfield  {journal} {\bibinfo  {journal} {arXiv e-prints}\ }\href {https://doi.org/10.48550/arXiv.2503.18877} {10.48550/arXiv.2503.18877} (\bibinfo {year} {2025})\BibitemShut {NoStop}%
\bibitem [{\citenamefont {Scalapino}(2012)}]{RevModPhys.84.1383}%
  \BibitemOpen
  \bibfield  {author} {\bibinfo {author} {\bibfnamefont {D.~J.}\ \bibnamefont {Scalapino}},\ }\bibfield  {title} {\bibinfo {title} {{A common thread: The pairing interaction for unconventional superconductors}},\ }\href {https://doi.org/10.1103/RevModPhys.84.1383} {\bibfield  {journal} {\bibinfo  {journal} {Rev. Mod. Phys.}\ }\textbf {\bibinfo {volume} {84}},\ \bibinfo {pages} {1383} (\bibinfo {year} {2012})}\BibitemShut {NoStop}%
\bibitem [{\citenamefont {{Wang}}\ \emph {et~al.}(2025{\natexlab{b}})\citenamefont {{Wang}}, \citenamefont {{Wang}}, \citenamefont {{Jiang}}, \citenamefont {{Hu}},\ and\ \citenamefont {{Zhang}}}]{jiangarxiv2025}%
  \BibitemOpen
  \bibfield  {author} {\bibinfo {author} {\bibfnamefont {Z.}~\bibnamefont {{Wang}}}, \bibinfo {author} {\bibfnamefont {Y.}~\bibnamefont {{Wang}}}, \bibinfo {author} {\bibfnamefont {K.}~\bibnamefont {{Jiang}}}, \bibinfo {author} {\bibfnamefont {J.}~\bibnamefont {{Hu}}},\ and\ \bibinfo {author} {\bibfnamefont {F.-C.}\ \bibnamefont {{Zhang}}},\ }\bibfield  {title} {\bibinfo {title} {{Discriminating Gap Symmetries of Superconducting $\mathrm{La}_{3}\mathrm{Ni}_{2}\mathrm{O}_{7}$}},\ }\href {https://doi.org/10.48550/arXiv.2512.12734} {\bibfield  {journal} {\bibinfo  {journal} {arXiv e-prints}\ ,\ \bibinfo {pages} {arXiv:2512.12734}} (\bibinfo {year} {2025}{\natexlab{b}})}\BibitemShut {NoStop}%
\bibitem [{\citenamefont {B{\"o}tzel}\ \emph {et~al.}(2025)\citenamefont {B{\"o}tzel}, \citenamefont {Lechermann}, \citenamefont {Shibauchi},\ and\ \citenamefont {Eremin}}]{Ilya_cp2025}%
  \BibitemOpen
  \bibfield  {author} {\bibinfo {author} {\bibfnamefont {S.}~\bibnamefont {B{\"o}tzel}}, \bibinfo {author} {\bibfnamefont {F.}~\bibnamefont {Lechermann}}, \bibinfo {author} {\bibfnamefont {T.}~\bibnamefont {Shibauchi}},\ and\ \bibinfo {author} {\bibfnamefont {I.~M.}\ \bibnamefont {Eremin}},\ }\bibfield  {title} {\bibinfo {title} {{Theory of potential impurity scattering in pressurized superconducting La$_3$Ni$_2$O$_7$}},\ }\href {https://doi.org/10.1038/s42005-025-02056-7} {\bibfield  {journal} {\bibinfo  {journal} {Communications Physics}\ }\textbf {\bibinfo {volume} {8}},\ \bibinfo {pages} {154} (\bibinfo {year} {2025})}\BibitemShut {NoStop}%
\bibitem [{\citenamefont {Huang}\ \emph {et~al.}(2025)\citenamefont {Huang}, \citenamefont {Wang},\ and\ \citenamefont {Zhou}}]{PhysRevB.111.174525}%
  \BibitemOpen
  \bibfield  {author} {\bibinfo {author} {\bibfnamefont {J.}~\bibnamefont {Huang}}, \bibinfo {author} {\bibfnamefont {Z.~D.}\ \bibnamefont {Wang}},\ and\ \bibinfo {author} {\bibfnamefont {T.}~\bibnamefont {Zhou}},\ }\bibfield  {title} {\bibinfo {title} {{Probing sign-changing order parameters via impurity states in unconventional superconductors: Implications for ${\mathrm{La}}_{3}{\mathrm{Ni}}_{2}{\mathrm{O}}_{7}$ superconductors with interlayer pairing}},\ }\href {https://doi.org/10.1103/PhysRevB.111.174525} {\bibfield  {journal} {\bibinfo  {journal} {Phys. Rev. B}\ }\textbf {\bibinfo {volume} {111}},\ \bibinfo {pages} {174525} (\bibinfo {year} {2025})}\BibitemShut {NoStop}%
\bibitem [{\citenamefont {B\"otzel}\ \emph {et~al.}(2024)\citenamefont {B\"otzel}, \citenamefont {Lechermann}, \citenamefont {Gondolf},\ and\ \citenamefont {Eremin}}]{PhysRevB.109.L180502}%
  \BibitemOpen
  \bibfield  {author} {\bibinfo {author} {\bibfnamefont {S.}~\bibnamefont {B\"otzel}}, \bibinfo {author} {\bibfnamefont {F.}~\bibnamefont {Lechermann}}, \bibinfo {author} {\bibfnamefont {J.}~\bibnamefont {Gondolf}},\ and\ \bibinfo {author} {\bibfnamefont {I.~M.}\ \bibnamefont {Eremin}},\ }\bibfield  {title} {\bibinfo {title} {{Theory of magnetic excitations in the multilayer nickelate superconductor ${\mathrm{La}}_{3}{\mathrm{Ni}}_{2}{\mathrm{O}}_{7}$}},\ }\href {https://doi.org/10.1103/PhysRevB.109.L180502} {\bibfield  {journal} {\bibinfo  {journal} {Phys. Rev. B}\ }\textbf {\bibinfo {volume} {109}},\ \bibinfo {pages} {L180502} (\bibinfo {year} {2024})}\BibitemShut {NoStop}%
\bibitem [{\citenamefont {Yang}(2025)}]{yang2025possible}%
  \BibitemOpen
  \bibfield  {author} {\bibinfo {author} {\bibfnamefont {Y.-f.}\ \bibnamefont {Yang}},\ }\bibfield  {title} {\bibinfo {title} {{Possible fano effect and suppression of Andreev reflection in ${\mathrm{La}}_{3}{\mathrm{Ni}}_{2}{\mathrm{O}}_{7}$}},\ }\href@noop {} {\bibfield  {journal} {\bibinfo  {journal} {Chinese Physics Letters}\ }\textbf {\bibinfo {volume} {42}},\ \bibinfo {pages} {017301} (\bibinfo {year} {2025})}\BibitemShut {NoStop}%
\bibitem [{\citenamefont {{Shen}}\ \emph {et~al.}(2025)\citenamefont {{Shen}}, \citenamefont {{Zhou}}, \citenamefont {{Miao}}, \citenamefont {{Li}}, \citenamefont {{Ou}}, \citenamefont {{Chen}}, \citenamefont {{Wang}}, \citenamefont {{Luan}}, \citenamefont {{Sun}}, \citenamefont {{Feng}}, \citenamefont {{Yong}}, \citenamefont {{Li}}, \citenamefont {{Xu}}, \citenamefont {{Lv}}, \citenamefont {{Nie}}, \citenamefont {{Wang}}, \citenamefont {{Huang}}, \citenamefont {{Sun}}, \citenamefont {{Xue}}, \citenamefont {{He}},\ and\ \citenamefont {{Chen}}}]{2025arXiv250217831S}%
  \BibitemOpen
  \bibfield  {author} {\bibinfo {author} {\bibfnamefont {J.}~\bibnamefont {{Shen}}}, \bibinfo {author} {\bibfnamefont {G.}~\bibnamefont {{Zhou}}}, \bibinfo {author} {\bibfnamefont {Y.}~\bibnamefont {{Miao}}}, \bibinfo {author} {\bibfnamefont {P.}~\bibnamefont {{Li}}}, \bibinfo {author} {\bibfnamefont {Z.}~\bibnamefont {{Ou}}}, \bibinfo {author} {\bibfnamefont {Y.}~\bibnamefont {{Chen}}}, \bibinfo {author} {\bibfnamefont {Z.}~\bibnamefont {{Wang}}}, \bibinfo {author} {\bibfnamefont {R.}~\bibnamefont {{Luan}}}, \bibinfo {author} {\bibfnamefont {H.}~\bibnamefont {{Sun}}}, \bibinfo {author} {\bibfnamefont {Z.}~\bibnamefont {{Feng}}}, \bibinfo {author} {\bibfnamefont {X.}~\bibnamefont {{Yong}}}, \bibinfo {author} {\bibfnamefont {Y.}~\bibnamefont {{Li}}}, \bibinfo {author} {\bibfnamefont {L.}~\bibnamefont {{Xu}}}, \bibinfo {author} {\bibfnamefont {W.}~\bibnamefont {{Lv}}}, \bibinfo {author} {\bibfnamefont {Z.}~\bibnamefont {{Nie}}}, \bibinfo {author} {\bibfnamefont {H.}~\bibnamefont {{Wang}}}, \bibinfo {author} {\bibfnamefont {H.}~\bibnamefont {{Huang}}}, \bibinfo {author} {\bibfnamefont {Y.-J.}\ \bibnamefont {{Sun}}}, \bibinfo {author} {\bibfnamefont {Q.-K.}\ \bibnamefont {{Xue}}}, \bibinfo {author} {\bibfnamefont {J.}~\bibnamefont {{He}}},\ and\ \bibinfo {author} {\bibfnamefont {Z.}~\bibnamefont {{Chen}}},\ }\bibfield  {title} {\bibinfo {title} {{Nodeless superconducting gap and electron-boson coupling in (La,Pr,Sm)$_{3}$Ni$_2$O$_7$ films}},\ }\bibfield  {journal} {\bibinfo  {journal} {arXiv e-prints}\ }\href {https://doi.org/10.48550/arXiv.2502.17831} {10.48550/arXiv.2502.17831} (\bibinfo {year} {2025})\BibitemShut {NoStop}%
\bibitem [{\citenamefont {{Sun}}\ \emph {et~al.}(2025)\citenamefont {{Sun}}, \citenamefont {{Jiang}}, \citenamefont {{Hao}}, \citenamefont {{Yan}}, \citenamefont {{Zhang}}, \citenamefont {{Wang}}, \citenamefont {{Yang}}, \citenamefont {{Sun}}, \citenamefont {{Liu}}, \citenamefont {{Ji}}, \citenamefont {{Gu}}, \citenamefont {{Zhou}}, \citenamefont {{Shen}}, \citenamefont {{Feng}},\ and\ \citenamefont {{Nie}}}]{2025arXiv250707409S}%
  \BibitemOpen
  \bibfield  {author} {\bibinfo {author} {\bibfnamefont {W.}~\bibnamefont {{Sun}}}, \bibinfo {author} {\bibfnamefont {Z.}~\bibnamefont {{Jiang}}}, \bibinfo {author} {\bibfnamefont {B.}~\bibnamefont {{Hao}}}, \bibinfo {author} {\bibfnamefont {S.}~\bibnamefont {{Yan}}}, \bibinfo {author} {\bibfnamefont {H.}~\bibnamefont {{Zhang}}}, \bibinfo {author} {\bibfnamefont {M.}~\bibnamefont {{Wang}}}, \bibinfo {author} {\bibfnamefont {Y.}~\bibnamefont {{Yang}}}, \bibinfo {author} {\bibfnamefont {H.}~\bibnamefont {{Sun}}}, \bibinfo {author} {\bibfnamefont {Z.}~\bibnamefont {{Liu}}}, \bibinfo {author} {\bibfnamefont {D.}~\bibnamefont {{Ji}}}, \bibinfo {author} {\bibfnamefont {Z.}~\bibnamefont {{Gu}}}, \bibinfo {author} {\bibfnamefont {J.}~\bibnamefont {{Zhou}}}, \bibinfo {author} {\bibfnamefont {D.}~\bibnamefont {{Shen}}}, \bibinfo {author} {\bibfnamefont {D.}~\bibnamefont {{Feng}}},\ and\ \bibinfo {author} {\bibfnamefont {Y.}~\bibnamefont {{Nie}}},\ }\bibfield  {title} {\bibinfo {title} {{Observation of superconductivity-induced leading-edge gap in Sr-doped $\mathrm{La}_{3}\mathrm{Ni}_{2}\mathrm{O}_{7}$ thin films}},\ }\href {https://doi.org/10.48550/arXiv.2507.07409} {\bibfield  {journal} {\bibinfo  {journal} {arXiv e-prints}\ ,\ \bibinfo {pages} {arXiv:2507.07409}} (\bibinfo {year} {2025})}\BibitemShut {NoStop}%
\bibitem [{\citenamefont {{Fan}}\ \emph {et~al.}(2025)\citenamefont {{Fan}}, \citenamefont {{Ou}}, \citenamefont {{Scholten}}, \citenamefont {{Li}}, \citenamefont {{Shang}}, \citenamefont {{Wang}}, \citenamefont {{Xu}}, \citenamefont {{Yang}}, \citenamefont {{Eremin}},\ and\ \citenamefont {{Wen}}}]{WenHH_STM2025}%
  \BibitemOpen
  \bibfield  {author} {\bibinfo {author} {\bibfnamefont {S.}~\bibnamefont {{Fan}}}, \bibinfo {author} {\bibfnamefont {M.}~\bibnamefont {{Ou}}}, \bibinfo {author} {\bibfnamefont {M.}~\bibnamefont {{Scholten}}}, \bibinfo {author} {\bibfnamefont {Q.}~\bibnamefont {{Li}}}, \bibinfo {author} {\bibfnamefont {Z.}~\bibnamefont {{Shang}}}, \bibinfo {author} {\bibfnamefont {Y.}~\bibnamefont {{Wang}}}, \bibinfo {author} {\bibfnamefont {J.}~\bibnamefont {{Xu}}}, \bibinfo {author} {\bibfnamefont {H.}~\bibnamefont {{Yang}}}, \bibinfo {author} {\bibfnamefont {I.~M.}\ \bibnamefont {{Eremin}}},\ and\ \bibinfo {author} {\bibfnamefont {H.-H.}\ \bibnamefont {{Wen}}},\ }\bibfield  {title} {\bibinfo {title} {{Superconducting gap structure and bosonic mode in La$_2$PrNi$_2$O$_7$ thin films at ambient pressure}},\ }\bibfield  {journal} {\bibinfo  {journal} {arXiv e-prints}\ }\href {https://doi.org/10.48550/arXiv.2506.01788} {10.48550/arXiv.2506.01788} (\bibinfo {year} {2025})\BibitemShut {NoStop}%
\bibitem [{\citenamefont {Hoffman}\ \emph {et~al.}(2002)\citenamefont {Hoffman}, \citenamefont {McElroy}, \citenamefont {Lee}, \citenamefont {Lang}, \citenamefont {Eisaki}, \citenamefont {Uchida},\ and\ \citenamefont {Davis}}]{doi:10.1126/science.1072640}%
  \BibitemOpen
  \bibfield  {author} {\bibinfo {author} {\bibfnamefont {J.~E.}\ \bibnamefont {Hoffman}}, \bibinfo {author} {\bibfnamefont {K.}~\bibnamefont {McElroy}}, \bibinfo {author} {\bibfnamefont {D.-H.}\ \bibnamefont {Lee}}, \bibinfo {author} {\bibfnamefont {K.~M.}\ \bibnamefont {Lang}}, \bibinfo {author} {\bibfnamefont {H.}~\bibnamefont {Eisaki}}, \bibinfo {author} {\bibfnamefont {S.}~\bibnamefont {Uchida}},\ and\ \bibinfo {author} {\bibfnamefont {J.~C.}\ \bibnamefont {Davis}},\ }\bibfield  {title} {\bibinfo {title} {{Imaging Quasiparticle Interference in Bi$_2$Sr$_2$CaCu$_2$O$_{8+\delta}$}},\ }\href {https://doi.org/10.1126/science.1072640} {\bibfield  {journal} {\bibinfo  {journal} {Science}\ }\textbf {\bibinfo {volume} {297}},\ \bibinfo {pages} {1148} (\bibinfo {year} {2002})},\ \Eprint {https://arxiv.org/abs/https://www.science.org/doi/pdf/10.1126/science.1072640} {https://www.science.org/doi/pdf/10.1126/science.1072640} \BibitemShut {NoStop}%
\bibitem [{\citenamefont {McElroy}\ \emph {et~al.}(2003)\citenamefont {McElroy}, \citenamefont {Simmonds}, \citenamefont {Hoffman}, \citenamefont {Lee}, \citenamefont {Orenstein}, \citenamefont {Eisaki}, \citenamefont {Uchida},\ and\ \citenamefont {Davis}}]{McElroy2003}%
  \BibitemOpen
  \bibfield  {author} {\bibinfo {author} {\bibfnamefont {K.}~\bibnamefont {McElroy}}, \bibinfo {author} {\bibfnamefont {R.~W.}\ \bibnamefont {Simmonds}}, \bibinfo {author} {\bibfnamefont {J.~E.}\ \bibnamefont {Hoffman}}, \bibinfo {author} {\bibfnamefont {D.~H.}\ \bibnamefont {Lee}}, \bibinfo {author} {\bibfnamefont {J.}~\bibnamefont {Orenstein}}, \bibinfo {author} {\bibfnamefont {H.}~\bibnamefont {Eisaki}}, \bibinfo {author} {\bibfnamefont {S.}~\bibnamefont {Uchida}},\ and\ \bibinfo {author} {\bibfnamefont {J.~C.}\ \bibnamefont {Davis}},\ }\bibfield  {title} {\bibinfo {title} {{Relating atomic-scale electronic phenomena to wave-like quasiparticle states in superconducting Bi$_2$Sr$_2$CaCu$_2$O$_{8+\delta}$}},\ }\href {https://doi.org/10.1038/nature01496} {\bibfield  {journal} {\bibinfo  {journal} {Nature}\ }\textbf {\bibinfo {volume} {422}},\ \bibinfo {pages} {592} (\bibinfo {year} {2003})}\BibitemShut {NoStop}%
\bibitem [{\citenamefont {Hanaguri}\ \emph {et~al.}(2009)\citenamefont {Hanaguri}, \citenamefont {Kohsaka}, \citenamefont {Ono}, \citenamefont {Maltseva}, \citenamefont {Coleman}, \citenamefont {Yamada}, \citenamefont {Azuma}, \citenamefont {Takano}, \citenamefont {Ohishi},\ and\ \citenamefont {Takagi}}]{hanaguri2009coherence}%
  \BibitemOpen
  \bibfield  {author} {\bibinfo {author} {\bibfnamefont {T.}~\bibnamefont {Hanaguri}}, \bibinfo {author} {\bibfnamefont {Y.}~\bibnamefont {Kohsaka}}, \bibinfo {author} {\bibfnamefont {M.}~\bibnamefont {Ono}}, \bibinfo {author} {\bibfnamefont {M.}~\bibnamefont {Maltseva}}, \bibinfo {author} {\bibfnamefont {P.}~\bibnamefont {Coleman}}, \bibinfo {author} {\bibfnamefont {I.}~\bibnamefont {Yamada}}, \bibinfo {author} {\bibfnamefont {M.}~\bibnamefont {Azuma}}, \bibinfo {author} {\bibfnamefont {M.}~\bibnamefont {Takano}}, \bibinfo {author} {\bibfnamefont {K.}~\bibnamefont {Ohishi}},\ and\ \bibinfo {author} {\bibfnamefont {H.}~\bibnamefont {Takagi}},\ }\bibfield  {title} {\bibinfo {title} {{Coherence factors in a high-$T_c$ cuprate probed by quasi-particle scattering off vortices}},\ }\href@noop {} {\bibfield  {journal} {\bibinfo  {journal} {Science}\ }\textbf {\bibinfo {volume} {323}},\ \bibinfo {pages} {923} (\bibinfo {year} {2009})}\BibitemShut {NoStop}%
\bibitem [{\citenamefont {Kohsaka}\ \emph {et~al.}(2008)\citenamefont {Kohsaka}, \citenamefont {Taylor}, \citenamefont {Wahl}, \citenamefont {Schmidt}, \citenamefont {Lee}, \citenamefont {Fujita}, \citenamefont {Alldredge}, \citenamefont {McElroy}, \citenamefont {Lee}, \citenamefont {Eisaki}, \citenamefont {Uchida}, \citenamefont {Lee},\ and\ \citenamefont {Davis}}]{Kohsaka2008}%
  \BibitemOpen
  \bibfield  {author} {\bibinfo {author} {\bibfnamefont {Y.}~\bibnamefont {Kohsaka}}, \bibinfo {author} {\bibfnamefont {C.}~\bibnamefont {Taylor}}, \bibinfo {author} {\bibfnamefont {P.}~\bibnamefont {Wahl}}, \bibinfo {author} {\bibfnamefont {A.}~\bibnamefont {Schmidt}}, \bibinfo {author} {\bibfnamefont {J.}~\bibnamefont {Lee}}, \bibinfo {author} {\bibfnamefont {K.}~\bibnamefont {Fujita}}, \bibinfo {author} {\bibfnamefont {J.~W.}\ \bibnamefont {Alldredge}}, \bibinfo {author} {\bibfnamefont {K.}~\bibnamefont {McElroy}}, \bibinfo {author} {\bibfnamefont {J.}~\bibnamefont {Lee}}, \bibinfo {author} {\bibfnamefont {H.}~\bibnamefont {Eisaki}}, \bibinfo {author} {\bibfnamefont {S.}~\bibnamefont {Uchida}}, \bibinfo {author} {\bibfnamefont {D.~H.}\ \bibnamefont {Lee}},\ and\ \bibinfo {author} {\bibfnamefont {J.~C.}\ \bibnamefont {Davis}},\ }\bibfield  {title} {\bibinfo {title} {{How Cooper pairs vanish approaching the Mott insulator in Bi$_2$Sr$_2$CaCu$_2$O$_{8+\delta}$}},\ }\href {https://doi.org/10.1038/nature07243} {\bibfield  {journal} {\bibinfo  {journal} {Nature}\ }\textbf {\bibinfo {volume} {454}},\ \bibinfo {pages} {1072} (\bibinfo {year} {2008})}\BibitemShut {NoStop}%
\bibitem [{\citenamefont {Allan}\ \emph {et~al.}(2013)\citenamefont {Allan}, \citenamefont {Chuang}, \citenamefont {Massee}, \citenamefont {Xie}, \citenamefont {Ni}, \citenamefont {Bud'ko}, \citenamefont {Boebinger}, \citenamefont {Wang}, \citenamefont {Dessau}, \citenamefont {Canfield}, \citenamefont {Golden},\ and\ \citenamefont {Davis}}]{Allan2013}%
  \BibitemOpen
  \bibfield  {author} {\bibinfo {author} {\bibfnamefont {M.~P.}\ \bibnamefont {Allan}}, \bibinfo {author} {\bibfnamefont {T.-M.}\ \bibnamefont {Chuang}}, \bibinfo {author} {\bibfnamefont {F.}~\bibnamefont {Massee}}, \bibinfo {author} {\bibfnamefont {Y.}~\bibnamefont {Xie}}, \bibinfo {author} {\bibfnamefont {N.}~\bibnamefont {Ni}}, \bibinfo {author} {\bibfnamefont {S.~L.}\ \bibnamefont {Bud'ko}}, \bibinfo {author} {\bibfnamefont {G.~S.}\ \bibnamefont {Boebinger}}, \bibinfo {author} {\bibfnamefont {Q.}~\bibnamefont {Wang}}, \bibinfo {author} {\bibfnamefont {D.~S.}\ \bibnamefont {Dessau}}, \bibinfo {author} {\bibfnamefont {P.~C.}\ \bibnamefont {Canfield}}, \bibinfo {author} {\bibfnamefont {M.~S.}\ \bibnamefont {Golden}},\ and\ \bibinfo {author} {\bibfnamefont {J.~C.}\ \bibnamefont {Davis}},\ }\bibfield  {title} {\bibinfo {title} {{Anisotropic impurity states, quasiparticle scattering and nematic transport in underdoped Ca(Fe$_{1-x}$Co$_{x}$)$_2$As$_2$}},\ }\href {https://doi.org/10.1038/nphys2544} {\bibfield  {journal} {\bibinfo  {journal} {Nature Physics}\ }\textbf {\bibinfo {volume} {9}},\ \bibinfo {pages} {220} (\bibinfo {year} {2013})}\BibitemShut {NoStop}%
\bibitem [{\citenamefont {Hanaguri}\ \emph {et~al.}(2010)\citenamefont {Hanaguri}, \citenamefont {Niitaka}, \citenamefont {Kuroki},\ and\ \citenamefont {Takagi}}]{hanaguri2010unconventional}%
  \BibitemOpen
  \bibfield  {author} {\bibinfo {author} {\bibfnamefont {T.}~\bibnamefont {Hanaguri}}, \bibinfo {author} {\bibfnamefont {S.}~\bibnamefont {Niitaka}}, \bibinfo {author} {\bibfnamefont {K.}~\bibnamefont {Kuroki}},\ and\ \bibinfo {author} {\bibfnamefont {H.}~\bibnamefont {Takagi}},\ }\bibfield  {title} {\bibinfo {title} {{Unconventional $s$-wave superconductivity in Fe(Se,Te)}},\ }\href@noop {} {\bibfield  {journal} {\bibinfo  {journal} {Science}\ }\textbf {\bibinfo {volume} {328}},\ \bibinfo {pages} {474} (\bibinfo {year} {2010})}\BibitemShut {NoStop}%
\bibitem [{\citenamefont {Le}\ \emph {et~al.}(2025)\citenamefont {Le}, \citenamefont {Zhan}, \citenamefont {Wu},\ and\ \citenamefont {Hu}}]{le2025landscape}%
  \BibitemOpen
  \bibfield  {author} {\bibinfo {author} {\bibfnamefont {C.}~\bibnamefont {Le}}, \bibinfo {author} {\bibfnamefont {J.}~\bibnamefont {Zhan}}, \bibinfo {author} {\bibfnamefont {X.}~\bibnamefont {Wu}},\ and\ \bibinfo {author} {\bibfnamefont {J.}~\bibnamefont {Hu}},\ }\bibfield  {title} {\bibinfo {title} {{Landscape of Correlated Orders in Strained Bilayer Nickelate Thin Films}},\ }\href {https://arxiv.org/abs/2501.14665} {\bibfield  {journal} {\bibinfo  {journal} {arXiv preprint arXiv:2501.14665}\ } (\bibinfo {year} {2025})}\BibitemShut {NoStop}%
\bibitem [{\citenamefont {Balatsky}\ \emph {et~al.}(2006)\citenamefont {Balatsky}, \citenamefont {Vekhter},\ and\ \citenamefont {Zhu}}]{RevModPhys.78.373}%
  \BibitemOpen
  \bibfield  {author} {\bibinfo {author} {\bibfnamefont {A.~V.}\ \bibnamefont {Balatsky}}, \bibinfo {author} {\bibfnamefont {I.}~\bibnamefont {Vekhter}},\ and\ \bibinfo {author} {\bibfnamefont {J.-X.}\ \bibnamefont {Zhu}},\ }\bibfield  {title} {\bibinfo {title} {Impurity-induced states in conventional and unconventional superconductors},\ }\href {https://doi.org/10.1103/RevModPhys.78.373} {\bibfield  {journal} {\bibinfo  {journal} {Rev. Mod. Phys.}\ }\textbf {\bibinfo {volume} {78}},\ \bibinfo {pages} {373} (\bibinfo {year} {2006})}\BibitemShut {NoStop}%
\bibitem [{\citenamefont {Fang}\ \emph {et~al.}(2013)\citenamefont {Fang}, \citenamefont {Gilbert}, \citenamefont {Xu}, \citenamefont {Bernevig},\ and\ \citenamefont {Hasan}}]{PhysRevB.88.125141}%
  \BibitemOpen
  \bibfield  {author} {\bibinfo {author} {\bibfnamefont {C.}~\bibnamefont {Fang}}, \bibinfo {author} {\bibfnamefont {M.~J.}\ \bibnamefont {Gilbert}}, \bibinfo {author} {\bibfnamefont {S.-Y.}\ \bibnamefont {Xu}}, \bibinfo {author} {\bibfnamefont {B.~A.}\ \bibnamefont {Bernevig}},\ and\ \bibinfo {author} {\bibfnamefont {M.~Z.}\ \bibnamefont {Hasan}},\ }\bibfield  {title} {\bibinfo {title} {Theory of quasiparticle interference in mirror-symmetric two-dimensional systems and its application to surface states of topological crystalline insulators},\ }\href {https://doi.org/10.1103/PhysRevB.88.125141} {\bibfield  {journal} {\bibinfo  {journal} {Phys. Rev. B}\ }\textbf {\bibinfo {volume} {88}},\ \bibinfo {pages} {125141} (\bibinfo {year} {2013})}\BibitemShut {NoStop}%
\bibitem [{\citenamefont {Zhan}\ \emph {et~al.}(2025)\citenamefont {Zhan}, \citenamefont {Gu}, \citenamefont {Wu},\ and\ \citenamefont {Hu}}]{PhysRevLett.134.136002}%
  \BibitemOpen
  \bibfield  {author} {\bibinfo {author} {\bibfnamefont {J.}~\bibnamefont {Zhan}}, \bibinfo {author} {\bibfnamefont {Y.}~\bibnamefont {Gu}}, \bibinfo {author} {\bibfnamefont {X.}~\bibnamefont {Wu}},\ and\ \bibinfo {author} {\bibfnamefont {J.}~\bibnamefont {Hu}},\ }\bibfield  {title} {\bibinfo {title} {{Cooperation between Electron-Phonon Coupling and Electronic Interaction in Bilayer Nickelates ${\mathrm{La}}_{3}{\mathrm{Ni}}_{2}{\mathrm{O}}_{7}$}},\ }\href {https://doi.org/10.1103/PhysRevLett.134.136002} {\bibfield  {journal} {\bibinfo  {journal} {Phys. Rev. Lett.}\ }\textbf {\bibinfo {volume} {134}},\ \bibinfo {pages} {136002} (\bibinfo {year} {2025})}\BibitemShut {NoStop}%
\bibitem [{\citenamefont {Sakakibara}\ \emph {et~al.}(2024{\natexlab{b}})\citenamefont {Sakakibara}, \citenamefont {Kitamine}, \citenamefont {Ochi},\ and\ \citenamefont {Kuroki}}]{PhysRevLett.132.106002}%
  \BibitemOpen
  \bibfield  {author} {\bibinfo {author} {\bibfnamefont {H.}~\bibnamefont {Sakakibara}}, \bibinfo {author} {\bibfnamefont {N.}~\bibnamefont {Kitamine}}, \bibinfo {author} {\bibfnamefont {M.}~\bibnamefont {Ochi}},\ and\ \bibinfo {author} {\bibfnamefont {K.}~\bibnamefont {Kuroki}},\ }\bibfield  {title} {\bibinfo {title} {{Possible High ${T}_{c}$ Superconductivity in ${\mathrm{La}}_{3}{\mathrm{Ni}}_{2}{\mathrm{O}}_{7}$ under High Pressure through Manifestation of a Nearly Half-Filled Bilayer Hubbard Model}},\ }\href {https://doi.org/10.1103/PhysRevLett.132.106002} {\bibfield  {journal} {\bibinfo  {journal} {Phys. Rev. Lett.}\ }\textbf {\bibinfo {volume} {132}},\ \bibinfo {pages} {106002} (\bibinfo {year} {2024}{\natexlab{b}})}\BibitemShut {NoStop}%
\bibitem [{\citenamefont {Gu}\ \emph {et~al.}(2025{\natexlab{b}})\citenamefont {Gu}, \citenamefont {Le}, \citenamefont {Yang}, \citenamefont {Wu},\ and\ \citenamefont {Hu}}]{PhysRevB.111.174506}%
  \BibitemOpen
  \bibfield  {author} {\bibinfo {author} {\bibfnamefont {Y.}~\bibnamefont {Gu}}, \bibinfo {author} {\bibfnamefont {C.}~\bibnamefont {Le}}, \bibinfo {author} {\bibfnamefont {Z.}~\bibnamefont {Yang}}, \bibinfo {author} {\bibfnamefont {X.}~\bibnamefont {Wu}},\ and\ \bibinfo {author} {\bibfnamefont {J.}~\bibnamefont {Hu}},\ }\bibfield  {title} {\bibinfo {title} {{Effective model and pairing tendency in the bilayer Ni-based superconductor ${\mathrm{La}}_{3}{\mathrm{Ni}}_{2}{\mathrm{O}}_{7}$}},\ }\href {https://doi.org/10.1103/PhysRevB.111.174506} {\bibfield  {journal} {\bibinfo  {journal} {Phys. Rev. B}\ }\textbf {\bibinfo {volume} {111}},\ \bibinfo {pages} {174506} (\bibinfo {year} {2025}{\natexlab{b}})}\BibitemShut {NoStop}%
\bibitem [{\citenamefont {Yang}\ \emph {et~al.}(2023{\natexlab{c}})\citenamefont {Yang}, \citenamefont {Wang},\ and\ \citenamefont {Wang}}]{PhysRevB.108.L140505}%
  \BibitemOpen
  \bibfield  {author} {\bibinfo {author} {\bibfnamefont {Q.-G.}\ \bibnamefont {Yang}}, \bibinfo {author} {\bibfnamefont {D.}~\bibnamefont {Wang}},\ and\ \bibinfo {author} {\bibfnamefont {Q.-H.}\ \bibnamefont {Wang}},\ }\bibfield  {title} {\bibinfo {title} {{Possible ${s}_{\ifmmode\pm\else\textpm\fi{}}$-wave superconductivity in ${\mathrm{La}}_{3}{\mathrm{Ni}}_{2}{\mathrm{O}}_{7}$}},\ }\href {https://doi.org/10.1103/PhysRevB.108.L140505} {\bibfield  {journal} {\bibinfo  {journal} {Phys. Rev. B}\ }\textbf {\bibinfo {volume} {108}},\ \bibinfo {pages} {L140505} (\bibinfo {year} {2023}{\natexlab{c}})}\BibitemShut {NoStop}%
\bibitem [{\citenamefont {Liu}\ \emph {et~al.}(2023{\natexlab{b}})\citenamefont {Liu}, \citenamefont {Mei}, \citenamefont {Ye}, \citenamefont {Chen},\ and\ \citenamefont {Yang}}]{PhysRevLett.131.236002}%
  \BibitemOpen
  \bibfield  {author} {\bibinfo {author} {\bibfnamefont {Y.-B.}\ \bibnamefont {Liu}}, \bibinfo {author} {\bibfnamefont {J.-W.}\ \bibnamefont {Mei}}, \bibinfo {author} {\bibfnamefont {F.}~\bibnamefont {Ye}}, \bibinfo {author} {\bibfnamefont {W.-Q.}\ \bibnamefont {Chen}},\ and\ \bibinfo {author} {\bibfnamefont {F.}~\bibnamefont {Yang}},\ }\bibfield  {title} {\bibinfo {title} {{${s}^{\ifmmode\pm\else\textpm\fi{}}$-Wave Pairing and the Destructive Role of Apical-Oxygen Deficiencies in ${\mathrm{La}}_{3}{\mathrm{Ni}}_{2}{\mathrm{O}}_{7}$ under Pressure}},\ }\href {https://doi.org/10.1103/PhysRevLett.131.236002} {\bibfield  {journal} {\bibinfo  {journal} {Phys. Rev. Lett.}\ }\textbf {\bibinfo {volume} {131}},\ \bibinfo {pages} {236002} (\bibinfo {year} {2023}{\natexlab{b}})}\BibitemShut {NoStop}%
\bibitem [{\citenamefont {Zhang}\ \emph {et~al.}(2024{\natexlab{b}})\citenamefont {Zhang}, \citenamefont {Lin}, \citenamefont {Moreo}, \citenamefont {Maier},\ and\ \citenamefont {Dagotto}}]{10.1038/s41467-024-46622-z}%
  \BibitemOpen
  \bibfield  {author} {\bibinfo {author} {\bibfnamefont {Y.}~\bibnamefont {Zhang}}, \bibinfo {author} {\bibfnamefont {L.-F.}\ \bibnamefont {Lin}}, \bibinfo {author} {\bibfnamefont {A.}~\bibnamefont {Moreo}}, \bibinfo {author} {\bibfnamefont {T.~A.}\ \bibnamefont {Maier}},\ and\ \bibinfo {author} {\bibfnamefont {E.}~\bibnamefont {Dagotto}},\ }\bibfield  {title} {\bibinfo {title} {{Structural phase transition, s$_\pm$-wave pairing, and magnetic stripe order in bilayered superconductor La$_3$Ni$_2$O$_7$ under pressure}},\ }\href {https://doi.org/10.1038/s41467-024-46622-z} {\bibfield  {journal} {\bibinfo  {journal} {Nature Communications}\ }\textbf {\bibinfo {volume} {15}},\ \bibinfo {pages} {2470} (\bibinfo {year} {2024}{\natexlab{b}})}\BibitemShut {NoStop}%
\bibitem [{\citenamefont {Xia}\ \emph {et~al.}(2025{\natexlab{b}})\citenamefont {Xia}, \citenamefont {Liu}, \citenamefont {Zhou},\ and\ \citenamefont {Chen}}]{10.1038/s41467-025-56206-0}%
  \BibitemOpen
  \bibfield  {author} {\bibinfo {author} {\bibfnamefont {C.}~\bibnamefont {Xia}}, \bibinfo {author} {\bibfnamefont {H.}~\bibnamefont {Liu}}, \bibinfo {author} {\bibfnamefont {S.}~\bibnamefont {Zhou}},\ and\ \bibinfo {author} {\bibfnamefont {H.}~\bibnamefont {Chen}},\ }\bibfield  {title} {\bibinfo {title} {{Sensitive dependence of pairing symmetry on Ni-eg crystal field splitting in the nickelate superconductor ${\mathrm{La}}_{3}{\mathrm{Ni}}_{2}{\mathrm{O}}_{7}$}},\ }\href {https://doi.org/10.1038/s41467-025-56206-0} {\bibfield  {journal} {\bibinfo  {journal} {Nature Communications}\ }\textbf {\bibinfo {volume} {16}},\ \bibinfo {pages} {1054} (\bibinfo {year} {2025}{\natexlab{b}})}\BibitemShut {NoStop}%
\bibitem [{\citenamefont {Zhang}\ \emph {et~al.}(2023{\natexlab{c}})\citenamefont {Zhang}, \citenamefont {Lin}, \citenamefont {Moreo}, \citenamefont {Maier},\ and\ \citenamefont {Dagotto}}]{PhysRevB.108.165141}%
  \BibitemOpen
  \bibfield  {author} {\bibinfo {author} {\bibfnamefont {Y.}~\bibnamefont {Zhang}}, \bibinfo {author} {\bibfnamefont {L.-F.}\ \bibnamefont {Lin}}, \bibinfo {author} {\bibfnamefont {A.}~\bibnamefont {Moreo}}, \bibinfo {author} {\bibfnamefont {T.~A.}\ \bibnamefont {Maier}},\ and\ \bibinfo {author} {\bibfnamefont {E.}~\bibnamefont {Dagotto}},\ }\bibfield  {title} {\bibinfo {title} {{Trends in electronic structures and ${s}_{\ifmmode\pm\else\textpm\fi{}}$-wave pairing for the rare-earth series in bilayer nickelate superconductor ${R}_{3}{\mathrm{Ni}}_{2}{\mathrm{O}}_{7}$}},\ }\href {https://doi.org/10.1103/PhysRevB.108.165141} {\bibfield  {journal} {\bibinfo  {journal} {Phys. Rev. B}\ }\textbf {\bibinfo {volume} {108}},\ \bibinfo {pages} {165141} (\bibinfo {year} {2023}{\natexlab{c}})}\BibitemShut {NoStop}%
\bibitem [{\citenamefont {Qu}\ \emph {et~al.}(2024{\natexlab{b}})\citenamefont {Qu}, \citenamefont {Qu}, \citenamefont {Chen}, \citenamefont {Wu}, \citenamefont {Yang}, \citenamefont {Li},\ and\ \citenamefont {Su}}]{PhysRevLett.132.036502}%
  \BibitemOpen
  \bibfield  {author} {\bibinfo {author} {\bibfnamefont {X.-Z.}\ \bibnamefont {Qu}}, \bibinfo {author} {\bibfnamefont {D.-W.}\ \bibnamefont {Qu}}, \bibinfo {author} {\bibfnamefont {J.}~\bibnamefont {Chen}}, \bibinfo {author} {\bibfnamefont {C.}~\bibnamefont {Wu}}, \bibinfo {author} {\bibfnamefont {F.}~\bibnamefont {Yang}}, \bibinfo {author} {\bibfnamefont {W.}~\bibnamefont {Li}},\ and\ \bibinfo {author} {\bibfnamefont {G.}~\bibnamefont {Su}},\ }\bibfield  {title} {\bibinfo {title} {{Bilayer ${t\text{\ensuremath{-}}J\text{\ensuremath{-}}J}_{\ensuremath{\perp}}$ Model and Magnetically Mediated Pairing in the Pressurized Nickelate ${\mathrm{La}}_{3}{\mathrm{Ni}}_{2}{\mathrm{O}}_{7}$}},\ }\href {https://doi.org/10.1103/PhysRevLett.132.036502} {\bibfield  {journal} {\bibinfo  {journal} {Phys. Rev. Lett.}\ }\textbf {\bibinfo {volume} {132}},\ \bibinfo {pages} {036502} (\bibinfo {year} {2024}{\natexlab{b}})}\BibitemShut {NoStop}%
\bibitem [{\citenamefont {Tian}\ \emph {et~al.}(2024{\natexlab{b}})\citenamefont {Tian}, \citenamefont {Chen}, \citenamefont {Wang}, \citenamefont {He},\ and\ \citenamefont {Lu}}]{PhysRevB.109.165154}%
  \BibitemOpen
  \bibfield  {author} {\bibinfo {author} {\bibfnamefont {Y.-H.}\ \bibnamefont {Tian}}, \bibinfo {author} {\bibfnamefont {Y.}~\bibnamefont {Chen}}, \bibinfo {author} {\bibfnamefont {J.-M.}\ \bibnamefont {Wang}}, \bibinfo {author} {\bibfnamefont {R.-Q.}\ \bibnamefont {He}},\ and\ \bibinfo {author} {\bibfnamefont {Z.-Y.}\ \bibnamefont {Lu}},\ }\bibfield  {title} {\bibinfo {title} {{Correlation effects and concomitant two-orbital ${s}_{\ifmmode\pm\else\textpm\fi{}}$-wave superconductivity in ${\mathrm{La}}_{3}{\mathrm{Ni}}_{2}{\mathrm{O}}_{7}$ under high pressure}},\ }\href {https://doi.org/10.1103/PhysRevB.109.165154} {\bibfield  {journal} {\bibinfo  {journal} {Phys. Rev. B}\ }\textbf {\bibinfo {volume} {109}},\ \bibinfo {pages} {165154} (\bibinfo {year} {2024}{\natexlab{b}})}\BibitemShut {NoStop}%
\bibitem [{\citenamefont {Liao}\ \emph {et~al.}(2023{\natexlab{b}})\citenamefont {Liao}, \citenamefont {Chen}, \citenamefont {Duan}, \citenamefont {Wang}, \citenamefont {Liu}, \citenamefont {Yu},\ and\ \citenamefont {Si}}]{PhysRevB.108.214522}%
  \BibitemOpen
  \bibfield  {author} {\bibinfo {author} {\bibfnamefont {Z.}~\bibnamefont {Liao}}, \bibinfo {author} {\bibfnamefont {L.}~\bibnamefont {Chen}}, \bibinfo {author} {\bibfnamefont {G.}~\bibnamefont {Duan}}, \bibinfo {author} {\bibfnamefont {Y.}~\bibnamefont {Wang}}, \bibinfo {author} {\bibfnamefont {C.}~\bibnamefont {Liu}}, \bibinfo {author} {\bibfnamefont {R.}~\bibnamefont {Yu}},\ and\ \bibinfo {author} {\bibfnamefont {Q.}~\bibnamefont {Si}},\ }\bibfield  {title} {\bibinfo {title} {{Electron correlations and superconductivity in ${\mathrm{La}}_{3}{\mathrm{Ni}}_{2}{\mathrm{O}}_{7}$ under pressure tuning}},\ }\href {https://doi.org/10.1103/PhysRevB.108.214522} {\bibfield  {journal} {\bibinfo  {journal} {Phys. Rev. B}\ }\textbf {\bibinfo {volume} {108}},\ \bibinfo {pages} {214522} (\bibinfo {year} {2023}{\natexlab{b}})}\BibitemShut {NoStop}%
\bibitem [{\citenamefont {Lu}\ \emph {et~al.}(2024{\natexlab{b}})\citenamefont {Lu}, \citenamefont {Pan}, \citenamefont {Yang},\ and\ \citenamefont {Wu}}]{PhysRevLett.132.146002}%
  \BibitemOpen
  \bibfield  {author} {\bibinfo {author} {\bibfnamefont {C.}~\bibnamefont {Lu}}, \bibinfo {author} {\bibfnamefont {Z.}~\bibnamefont {Pan}}, \bibinfo {author} {\bibfnamefont {F.}~\bibnamefont {Yang}},\ and\ \bibinfo {author} {\bibfnamefont {C.}~\bibnamefont {Wu}},\ }\bibfield  {title} {\bibinfo {title} {{Interlayer-Coupling-Driven High-Temperature Superconductivity in ${\mathrm{La}}_{3}{\mathrm{Ni}}_{2}{\mathrm{O}}_{7}$ under Pressure}},\ }\href {https://doi.org/10.1103/PhysRevLett.132.146002} {\bibfield  {journal} {\bibinfo  {journal} {Phys. Rev. Lett.}\ }\textbf {\bibinfo {volume} {132}},\ \bibinfo {pages} {146002} (\bibinfo {year} {2024}{\natexlab{b}})}\BibitemShut {NoStop}%
\bibitem [{\citenamefont {Oh}\ and\ \citenamefont {Zhang}(2023{\natexlab{b}})}]{PhysRevB.108.174511}%
  \BibitemOpen
  \bibfield  {author} {\bibinfo {author} {\bibfnamefont {H.}~\bibnamefont {Oh}}\ and\ \bibinfo {author} {\bibfnamefont {Y.-H.}\ \bibnamefont {Zhang}},\ }\bibfield  {title} {\bibinfo {title} {{Type-II $t\ensuremath{-}J$ model and shared superexchange coupling from Hund's rule in superconducting ${\mathrm{La}}_{3}{\mathrm{Ni}}_{2}{\mathrm{O}}_{7}$}},\ }\href {https://doi.org/10.1103/PhysRevB.108.174511} {\bibfield  {journal} {\bibinfo  {journal} {Phys. Rev. B}\ }\textbf {\bibinfo {volume} {108}},\ \bibinfo {pages} {174511} (\bibinfo {year} {2023}{\natexlab{b}})}\BibitemShut {NoStop}%
\bibitem [{\citenamefont {Yang}\ \emph {et~al.}(2023{\natexlab{d}})\citenamefont {Yang}, \citenamefont {Zhang},\ and\ \citenamefont {Zhang}}]{PhysRevB.108.L201108}%
  \BibitemOpen
  \bibfield  {author} {\bibinfo {author} {\bibfnamefont {Y.-f.}\ \bibnamefont {Yang}}, \bibinfo {author} {\bibfnamefont {G.-M.}\ \bibnamefont {Zhang}},\ and\ \bibinfo {author} {\bibfnamefont {F.-C.}\ \bibnamefont {Zhang}},\ }\bibfield  {title} {\bibinfo {title} {{Interlayer valence bonds and two-component theory for high-${T}_{c}$ superconductivity of ${\mathrm{La}}_{3}{\mathrm{Ni}}_{2}{\mathrm{O}}_{7}$ under pressure}},\ }\href {https://doi.org/10.1103/PhysRevB.108.L201108} {\bibfield  {journal} {\bibinfo  {journal} {Phys. Rev. B}\ }\textbf {\bibinfo {volume} {108}},\ \bibinfo {pages} {L201108} (\bibinfo {year} {2023}{\natexlab{d}})}\BibitemShut {NoStop}%
\bibitem [{\citenamefont {Fan}\ \emph {et~al.}(2024{\natexlab{b}})\citenamefont {Fan}, \citenamefont {Zhang}, \citenamefont {Zhan}, \citenamefont {Lv}, \citenamefont {Jiang}, \citenamefont {Normand},\ and\ \citenamefont {Xiang}}]{PhysRevB.110.024514}%
  \BibitemOpen
  \bibfield  {author} {\bibinfo {author} {\bibfnamefont {Z.}~\bibnamefont {Fan}}, \bibinfo {author} {\bibfnamefont {J.-F.}\ \bibnamefont {Zhang}}, \bibinfo {author} {\bibfnamefont {B.}~\bibnamefont {Zhan}}, \bibinfo {author} {\bibfnamefont {D.}~\bibnamefont {Lv}}, \bibinfo {author} {\bibfnamefont {X.-Y.}\ \bibnamefont {Jiang}}, \bibinfo {author} {\bibfnamefont {B.}~\bibnamefont {Normand}},\ and\ \bibinfo {author} {\bibfnamefont {T.}~\bibnamefont {Xiang}},\ }\bibfield  {title} {\bibinfo {title} {Superconductivity in nickelate and cuprate superconductors with strong bilayer coupling},\ }\href {https://doi.org/10.1103/PhysRevB.110.024514} {\bibfield  {journal} {\bibinfo  {journal} {Phys. Rev. B}\ }\textbf {\bibinfo {volume} {110}},\ \bibinfo {pages} {024514} (\bibinfo {year} {2024}{\natexlab{b}})}\BibitemShut {NoStop}%
\bibitem [{\citenamefont {Heier}\ \emph {et~al.}(2024)\citenamefont {Heier}, \citenamefont {Park},\ and\ \citenamefont {Savrasov}}]{PhysRevB.109.104508}%
  \BibitemOpen
  \bibfield  {author} {\bibinfo {author} {\bibfnamefont {G.}~\bibnamefont {Heier}}, \bibinfo {author} {\bibfnamefont {K.}~\bibnamefont {Park}},\ and\ \bibinfo {author} {\bibfnamefont {S.~Y.}\ \bibnamefont {Savrasov}},\ }\bibfield  {title} {\bibinfo {title} {{Competing ${d}_{xy}$ and ${s}_{\ifmmode\pm\else\textpm\fi{}}$ pairing symmetries in superconducting ${\mathrm{La}}_{3}{\mathrm{Ni}}_{2}{\mathrm{O}}_{7}$: $\mathrm{LDA}+\mathrm{FLEX}$ calculations}},\ }\href {https://doi.org/10.1103/PhysRevB.109.104508} {\bibfield  {journal} {\bibinfo  {journal} {Phys. Rev. B}\ }\textbf {\bibinfo {volume} {109}},\ \bibinfo {pages} {104508} (\bibinfo {year} {2024})}\BibitemShut {NoStop}%
\bibitem [{\citenamefont {Jiang}\ \emph {et~al.}(2024{\natexlab{d}})\citenamefont {Jiang}, \citenamefont {Hou}, \citenamefont {Fan}, \citenamefont {Lang},\ and\ \citenamefont {Ku}}]{PhysRevLett.132.126503}%
  \BibitemOpen
  \bibfield  {author} {\bibinfo {author} {\bibfnamefont {R.}~\bibnamefont {Jiang}}, \bibinfo {author} {\bibfnamefont {J.}~\bibnamefont {Hou}}, \bibinfo {author} {\bibfnamefont {Z.}~\bibnamefont {Fan}}, \bibinfo {author} {\bibfnamefont {Z.-J.}\ \bibnamefont {Lang}},\ and\ \bibinfo {author} {\bibfnamefont {W.}~\bibnamefont {Ku}},\ }\bibfield  {title} {\bibinfo {title} {{Pressure Driven Fractionalization of Ionic Spins Results in Cupratelike High-${T}_{c}$ Superconductivity in ${\mathrm{La}}_{3}{\mathrm{Ni}}_{2}{\mathrm{O}}_{7}$}},\ }\href {https://doi.org/10.1103/PhysRevLett.132.126503} {\bibfield  {journal} {\bibinfo  {journal} {Phys. Rev. Lett.}\ }\textbf {\bibinfo {volume} {132}},\ \bibinfo {pages} {126503} (\bibinfo {year} {2024}{\natexlab{d}})}\BibitemShut {NoStop}%
\bibitem [{\citenamefont {Maltseva}\ and\ \citenamefont {Coleman}(2009)}]{PhysRevB.80.144514}%
  \BibitemOpen
  \bibfield  {author} {\bibinfo {author} {\bibfnamefont {M.}~\bibnamefont {Maltseva}}\ and\ \bibinfo {author} {\bibfnamefont {P.}~\bibnamefont {Coleman}},\ }\bibfield  {title} {\bibinfo {title} {Model for nodal quasiparticle scattering in a disordered vortex lattice},\ }\href {https://doi.org/10.1103/PhysRevB.80.144514} {\bibfield  {journal} {\bibinfo  {journal} {Phys. Rev. B}\ }\textbf {\bibinfo {volume} {80}},\ \bibinfo {pages} {144514} (\bibinfo {year} {2009})}\BibitemShut {NoStop}%
\bibitem [{\citenamefont {Yamakawa}\ and\ \citenamefont {Kontani}(2015)}]{PhysRevB.92.045124}%
  \BibitemOpen
  \bibfield  {author} {\bibinfo {author} {\bibfnamefont {Y.}~\bibnamefont {Yamakawa}}\ and\ \bibinfo {author} {\bibfnamefont {H.}~\bibnamefont {Kontani}},\ }\bibfield  {title} {\bibinfo {title} {{Quasiparticle interference in Fe-based superconductors based on a five-orbital tight-binding model}},\ }\href {https://doi.org/10.1103/PhysRevB.92.045124} {\bibfield  {journal} {\bibinfo  {journal} {Phys. Rev. B}\ }\textbf {\bibinfo {volume} {92}},\ \bibinfo {pages} {045124} (\bibinfo {year} {2015})}\BibitemShut {NoStop}%
\bibitem [{\citenamefont {Wang}\ and\ \citenamefont {Lee}(2003)}]{PhysRevB.67.020511}%
  \BibitemOpen
  \bibfield  {author} {\bibinfo {author} {\bibfnamefont {Q.-H.}\ \bibnamefont {Wang}}\ and\ \bibinfo {author} {\bibfnamefont {D.-H.}\ \bibnamefont {Lee}},\ }\bibfield  {title} {\bibinfo {title} {{Quasiparticle scattering interference in high-temperature superconductors}},\ }\href {https://doi.org/10.1103/PhysRevB.67.020511} {\bibfield  {journal} {\bibinfo  {journal} {Phys. Rev. B}\ }\textbf {\bibinfo {volume} {67}},\ \bibinfo {pages} {020511} (\bibinfo {year} {2003})}\BibitemShut {NoStop}%
\bibitem [{\citenamefont {Sykora}\ and\ \citenamefont {Coleman}(2011)}]{PhysRevB.84.054501}%
  \BibitemOpen
  \bibfield  {author} {\bibinfo {author} {\bibfnamefont {S.}~\bibnamefont {Sykora}}\ and\ \bibinfo {author} {\bibfnamefont {P.}~\bibnamefont {Coleman}},\ }\bibfield  {title} {\bibinfo {title} {Quasiparticle interference in an iron-based superconductor},\ }\href {https://doi.org/10.1103/PhysRevB.84.054501} {\bibfield  {journal} {\bibinfo  {journal} {Phys. Rev. B}\ }\textbf {\bibinfo {volume} {84}},\ \bibinfo {pages} {054501} (\bibinfo {year} {2011})}\BibitemShut {NoStop}%
\bibitem [{\citenamefont {Hirschfeld}\ \emph {et~al.}(2015)\citenamefont {Hirschfeld}, \citenamefont {Altenfeld}, \citenamefont {Eremin},\ and\ \citenamefont {Mazin}}]{PhysRevB.92.184513}%
  \BibitemOpen
  \bibfield  {author} {\bibinfo {author} {\bibfnamefont {P.~J.}\ \bibnamefont {Hirschfeld}}, \bibinfo {author} {\bibfnamefont {D.}~\bibnamefont {Altenfeld}}, \bibinfo {author} {\bibfnamefont {I.}~\bibnamefont {Eremin}},\ and\ \bibinfo {author} {\bibfnamefont {I.~I.}\ \bibnamefont {Mazin}},\ }\bibfield  {title} {\bibinfo {title} {Robust determination of the superconducting gap sign structure via quasiparticle interference},\ }\href {https://doi.org/10.1103/PhysRevB.92.184513} {\bibfield  {journal} {\bibinfo  {journal} {Phys. Rev. B}\ }\textbf {\bibinfo {volume} {92}},\ \bibinfo {pages} {184513} (\bibinfo {year} {2015})}\BibitemShut {NoStop}%
\bibitem [{\citenamefont {Pereg-Barnea}\ and\ \citenamefont {Franz}(2008)}]{PhysRevB.78.020509}%
  \BibitemOpen
  \bibfield  {author} {\bibinfo {author} {\bibfnamefont {T.}~\bibnamefont {Pereg-Barnea}}\ and\ \bibinfo {author} {\bibfnamefont {M.}~\bibnamefont {Franz}},\ }\bibfield  {title} {\bibinfo {title} {Magnetic-field dependence of quasiparticle interference peaks in a $d$-wave superconductor with weak disorder},\ }\href {https://doi.org/10.1103/PhysRevB.78.020509} {\bibfield  {journal} {\bibinfo  {journal} {Phys. Rev. B}\ }\textbf {\bibinfo {volume} {78}},\ \bibinfo {pages} {020509} (\bibinfo {year} {2008})}\BibitemShut {NoStop}%
\bibitem [{\citenamefont {Chi}\ \emph {et~al.}(2017{\natexlab{a}})\citenamefont {Chi}, \citenamefont {Hardy}, \citenamefont {Liang}, \citenamefont {Dosanjh}, \citenamefont {Wahl}, \citenamefont {Burke},\ and\ \citenamefont {Bonn}}]{chi2017extracting}%
  \BibitemOpen
  \bibfield  {author} {\bibinfo {author} {\bibfnamefont {S.}~\bibnamefont {Chi}}, \bibinfo {author} {\bibfnamefont {W.}~\bibnamefont {Hardy}}, \bibinfo {author} {\bibfnamefont {R.}~\bibnamefont {Liang}}, \bibinfo {author} {\bibfnamefont {P.}~\bibnamefont {Dosanjh}}, \bibinfo {author} {\bibfnamefont {P.}~\bibnamefont {Wahl}}, \bibinfo {author} {\bibfnamefont {S.}~\bibnamefont {Burke}},\ and\ \bibinfo {author} {\bibfnamefont {D.}~\bibnamefont {Bonn}},\ }\bibfield  {title} {\bibinfo {title} {Extracting phase information about the superconducting order parameter from defect bound states},\ }\href@noop {} {\bibfield  {journal} {\bibinfo  {journal} {arXiv preprint arXiv:1710.09088}\ } (\bibinfo {year} {2017}{\natexlab{a}})}\BibitemShut {NoStop}%
\bibitem [{\citenamefont {Chi}\ \emph {et~al.}(2017{\natexlab{b}})\citenamefont {Chi}, \citenamefont {Hardy}, \citenamefont {Liang}, \citenamefont {Dosanjh}, \citenamefont {Wahl}, \citenamefont {Burke},\ and\ \citenamefont {Bonn}}]{chi2017determination}%
  \BibitemOpen
  \bibfield  {author} {\bibinfo {author} {\bibfnamefont {S.}~\bibnamefont {Chi}}, \bibinfo {author} {\bibfnamefont {W.}~\bibnamefont {Hardy}}, \bibinfo {author} {\bibfnamefont {R.}~\bibnamefont {Liang}}, \bibinfo {author} {\bibfnamefont {P.}~\bibnamefont {Dosanjh}}, \bibinfo {author} {\bibfnamefont {P.}~\bibnamefont {Wahl}}, \bibinfo {author} {\bibfnamefont {S.}~\bibnamefont {Burke}},\ and\ \bibinfo {author} {\bibfnamefont {D.}~\bibnamefont {Bonn}},\ }\bibfield  {title} {\bibinfo {title} {Determination of the superconducting order parameter from defect bound state quasiparticle interference},\ }\href@noop {} {\bibfield  {journal} {\bibinfo  {journal} {arXiv preprint arXiv:1710.09089}\ } (\bibinfo {year} {2017}{\natexlab{b}})}\BibitemShut {NoStop}%
\end{thebibliography}%
